\documentclass[prd,nofootinbib]{revtex4}

\usepackage{amsmath,amssymb}
\usepackage{rotating}
\usepackage{mathrsfs}
\usepackage{pstricks}
\usepackage{pst-func}
\usepackage{graphicx}
\usepackage{refcount} 
\usepackage{cases}
\usepackage{epsfig}
\usepackage{alltt}

\allowdisplaybreaks

\begin{document}

\title{Dr.~Bertlmann's Socks in the Quaternionic World of Ambidextral Reality}

\author{Joy Christian}

\email{jjc@alum.bu.edu}

\affiliation{Einstein Centre for Local-Realistic Physics, 15 Thackley End, Oxford OX2 6LB, United Kingdom}

\begin{abstract}
In this pedagogical paper, John S. Bell's amusing example of Dr.~Bertlmann's socks is reconsidered, first within a toy model of a two-dimensional one-sided world of a non-orientable M\"obius strip, and then within a real world of three-dimensional quaternionic sphere, $S^3$, which results from an addition of a single point to ${{\rm I\!R}^3}$ at infinity. In the quaternionic world, which happens to be the spatial part of a solution of Einstein's field equations of general relativity, the singlet correlations between a pair of\break entangled fermions can be understood as classically as those between Dr.~Bertlmann's colorful socks.
\end{abstract}

\maketitle

\parskip 7pt

\baselineskip 14.5pt

\section{Introduction}\label{sec:introduction}

The purpose of this paper is to illustrate the central ideas published in Refs.~\cite{Christian}, \cite{RSOS} and \cite{Christian2014} in a pedagogical manner. These ideas concern a local-realistic understanding of the origins and strengths of all possible quantum correlations \cite{disproof}.\break The ideal pedagogical device for this purpose is Bell's amusing example of Dr.~Bertlmann's colorful socks \cite{Bell-1987}. He writes:

\begin{quote}
The philosopher in the street, who has not suffered a course in quantum mechanics, is quite unimpressed by Einstein-Podolsky-Rosen correlations$^1$. He can point to many examples of similar correlations in everyday life. The case of Bertlmann's socks is often cited. Dr.~Bertlmann likes to wear two socks of different colours. Which colour he will have on a given foot on a given day is quite unpredictable. But when you see (Fig.~1)\break that the first sock is pink you can be already sure that the second sock will not be pink. Observation of the first, and experience of Bertlmann, gives immediate information about the second. There is no accounting for tastes, but apart from that there is no mystery here. And is not the EPR business just the same? \cite{Bell-1987}. 
\end{quote}

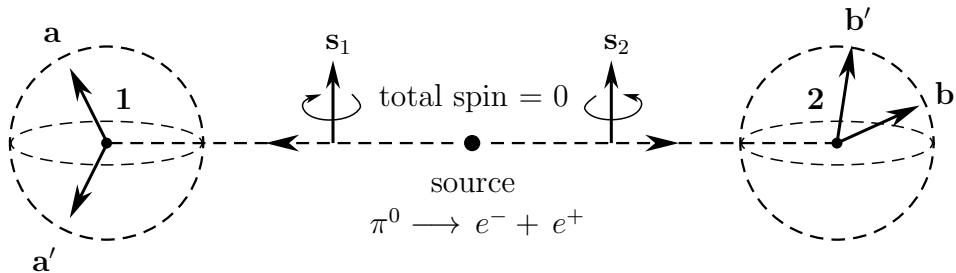
\begin{figure*}[t]
\hrule
\scalebox{1}{
\begin{pspicture}(1.2,-2.5)(4.2,2.5)

\psline[linewidth=0.1mm,dotsize=3pt 4]{*-}(-2.51,0)(-2.5,0)

\psline[linewidth=0.1mm,dotsize=3pt 4]{*-}(7.2,0)(7.15,0)

\psline[linewidth=0.4mm,arrowinset=0.3,arrowsize=3pt 3,arrowlength=2]{->}(-2.5,0)(-3,1)

\psline[linewidth=0.4mm,arrowinset=0.3,arrowsize=3pt 3,arrowlength=2]{->}(-2.5,0)(-3,-1)

\psline[linewidth=0.4mm,arrowinset=0.3,arrowsize=3pt 3,arrowlength=2]{->}(7.2,0)(8.3,0.5)

\psline[linewidth=0.4mm,arrowinset=0.3,arrowsize=3pt 3,arrowlength=2]{->}(7.2,0)(7.4,1.3)

\psline[linewidth=0.4mm,arrowinset=0.3,arrowsize=2pt 3,arrowlength=2]{->}(4.2,0)(4.2,1.1)

\psline[linewidth=0.4mm,arrowinset=0.3,arrowsize=2pt 3,arrowlength=2]{->}(0.5,0)(0.5,1.1)

\pscurve[linewidth=0.2mm,arrowinset=0.2,arrowsize=2pt 2,arrowlength=2]{->}(4.0,0.63)(3.85,0.45)(4.6,0.5)(4.35,0.65)

\put(4.1,1.25){{\large ${{\bf s}_2}$}}

\pscurve[linewidth=0.2mm,arrowinset=0.2,arrowsize=2pt 2,arrowlength=2]{<-}(0.35,0.65)(0.1,0.47)(0.86,0.47)(0.75,0.65)

\put(0.4,1.25){{\large ${{\bf s}_1}$}}

\put(-2.4,+0.45){{\large ${\bf 1}$}}

\put(6.8,+0.45){{\large ${\bf 2}$}}

\put(-3.35,1.35){{\large ${\bf a}$}}

\put(-3.5,-1.7){{\large ${\bf a'}$}}

\put(8.5,0.52){{\large ${\bf b}$}}

\put(7.3,1.5){{\large ${\bf b'}$}}

\put(1.8,-0.65){\large source}

\put(0.99,-1.2){\large ${\pi^0\longrightarrow\,e^{-}+\,e^{+}\,}$}

\put(1.11,0.5){\large total spin = 0}

\psline[linewidth=0.3mm,linestyle=dashed](-2.47,0)(2.1,0)

\psline[linewidth=0.4mm,arrowinset=0.3,arrowsize=3pt 3,arrowlength=2]{->}(-0.3,0)(-0.4,0)

\psline[linewidth=0.3mm,linestyle=dashed](2.6,0)(7.2,0)

\psline[linewidth=0.4mm,arrowinset=0.3,arrowsize=3pt 3,arrowlength=2]{->}(5.0,0)(5.1,0)

\psline[linewidth=0.1mm,dotsize=5pt 4]{*-}(2.35,0)(2.4,0)

\pscircle[linewidth=0.3mm,linestyle=dashed](7.2,0){1.3}

\psellipse[linewidth=0.2mm,linestyle=dashed](7.2,0)(1.28,0.3)

\pscircle[linewidth=0.3mm,linestyle=dashed](-2.51,0){1.3}

\psellipse[linewidth=0.2mm,linestyle=dashed](-2.51,0)(1.28,0.3)

\end{pspicture}}
\hrule
\caption{A spin-less neutral pion decays into an electron-positron pair (such a photon-less decay is rare but not impossible, and will suffice for our conceptual purposes here). Measurements of spin components on each separated fermion are performed at remote stations ${\mathbf{1}}$ and ${\mathbf{2}}$, providing binary outcomes along arbitrary directions such as ${\mathbf a}$ and ${\mathbf b}$, freely chosen by Alice and Bob.}
\vspace{0.25cm}
\hrule
\label{fig-1}
\end{figure*}

Later on in the discussion Bell points out that for the singlet correlations such a classical explanation works well so long as we restrict to perfect anti-correlation, which amounts to measuring the same component of spin at the two ends\break of an EPR type experimental set up (cf. Fig.~\ref{fig-1}). But as soon as we start measuring in non-parallel directions we get results that cannot be explained while respecting Einstein's notion of realism \cite{EPR}, which posits that spin components existed before they were measured \cite{Howard-2019}. To understand this ``mystery" better, let us use a variant of the above example discussed by Bell in his last paper, involving hand-gloves \cite{Bell-1990}, because it is more suitable for our purposes in this paper.

To that end, imagine you set out from home in a cold winter night and midway to your destination you reach out in your pockets for hand-gloves, only to find one of them. At that very moment you instantly know that you forgot the other glove at home. Not only do you know that, but you also instantly know --- and can predict with utmost confidence --- that the one you forgot at home will be seen to be left-handed if the one you pulled out from your pocket happens to be right-handed. Moreover, this will be true even if your home happens to be in the farthest corner of the Universe. You will have instant information about the handedness of the glove you forgot at home by simply looking at the one you just pulled out from your pocket. This is because the perfect anti-correlation between the handedness of the gloves preexisted regardless of anyone looking at them, and it has nothing to do with non-locality of any kind, because the handedness of one glove compared to the other had been pre-established before you set out from home. 

Let us now try to understand why this simple explanation appears to fail in the case of quantum mechanically predicted and experimentally verified singlet correlations \cite{Bell-1990}. Following the notations and terminology established in the first equation of Bell's famous paper \cite{Bell-1964}, let ${\mathscr{A}({\bf a},\,\lambda)=+1}$ or ${-1}$ represent the result of a measurement of a spin component along the detector direction ${\bf a}$ and ${\mathscr{B}({\bf b},\,\lambda)=+1}$ or ${-1}$ represent the result of a measurement of a spin component along the detector direction ${\bf b}$. Here ${\lambda}$ represents an initial or ``complete" state that has pre-established the harmony between these results, observed by the experimenters Alice and Bob at the two ends of an EPR-Bohm type experimental set up, as shown in Fig.~\ref{fig-1}. Now the conservation of spin angular momentum dictates that the total spin of the pair of fermions emerging from the source remains zero throughout the free evolution of the constituent spins. Therefore, if the result ${\mathscr{A}({\bf a},\,\lambda)}$ is observed to be equal to ${+1}$ (analogous to the glove being found right-handed), then the result ${\mathscr{B}({\bf b},\,\lambda)}$ would be necessarily equal to ${-1}$ (analogous to the glove being found left-handed). Consequently, the product of these results would necessarily satisfy ${\mathscr{A}({\bf a},\,\lambda)\,\mathscr{B}({\bf b},\,\lambda) = -1}$ for all ${\bf a}$ and ${\bf b}$. But quantum mechanics predicts that the product ${\mathscr{A}({\bf a},\,\lambda)\,\mathscr{B}({\bf b},\,\lambda)}$ can be equal to ${-1}$ only for a very special case when Bob has accidentally chosen a measurement direction that happens to be parallel to the one freely and independently chosen by Alice:
\begin{equation}
\mathscr{A}({\bf a},\,\lambda)\,\mathscr{B}({\bf b},\,\lambda) = -1 \;\;\text{when} \;\, {\bf b}=+{\bf a}. 
\end{equation}
Evidently, these perfect anti-correlation between the results of the spin measurements are entirely analogous to those between Dr.~Bertlmann's socks \cite{Bell-1987}. However, if Bob happens to choose a measurement direction that is {\it anti}-parallel to the one chosen by Alice, then quantum mechanics predicts a result that is difficult to understand in classical terms:
\begin{equation}
\mathscr{A}({\bf a},\,\lambda)\,\mathscr{B}({\bf b},\,\lambda) = +1 \;\;\text{when} \;\, {\bf b}=-{\bf a}. \label{2}
\end{equation}
This prediction dictates that if we happen to find a right-handed glove [{\it i.e.}, ${\mathscr{A}({\bf a},\,\lambda)=+1}$] in our pocket, then the one forgotten at home would also be right-handed [{\it i.e.}, ${\mathscr{B}({\bf b},\,\lambda)=+1}$]. And if we happen to find a left-handed glove [{\it i.e.}, ${\mathscr{A}({\bf a},\,\lambda)=-1}$] in our pocket, then the one forgotten at home would also be left-handed [{\it i.e.}, ${\mathscr{B}({\bf b},\,\lambda)=-1}$]. That, of course, seems completely at odds with our classical intuition of gloves having preexisting handedness. Moreover, for intermediate cases with ${{\bf b}\not={\bf a}}$ quantum mechanics predicts ${\mathscr{A}({\bf a},\,\lambda)\,\mathscr{B}({\bf b},\,\lambda) = +1}$ or ${-1}$, so that, on the average, 
\begin{equation}
\Bigl\langle\,\mathscr{A}({\bf a},\,\lambda)\,\mathscr{B}({\bf b},\,\lambda)\,\Bigr\rangle = -{\bf a}\cdot{\bf b},\;\;\;\text{together with}\;\;\; \Bigl\langle\,\mathscr{A}({\bf a},\,\lambda)\,\Bigr\rangle=0 \;\;\;\text{and}\;\;\;\Bigl\langle\,\mathscr{B}({\bf b},\,\lambda)\,\Bigr\rangle = 0.   \label{4}
\end{equation}
These predictions clearly require sign-flips from ${\mathscr{A}({\bf a},\,\lambda)\,\mathscr{B}({\bf b},\,\lambda) = -1}$ to ${\mathscr{A}({\bf a},\,\lambda)\,\mathscr{B}({\bf b},\,\lambda) = +1}$ for at least some of the choices of measurement directions made by Alice and Bob so that the average of the product can result in ${-{\bf a}\cdot{\bf b}}$. That would be mathematically impossible if the value of the product remained equal to ${-1}$ for all choices of ${\bf a}$ and ${\bf b}$, as in Dr.~Bertlmann's socks type correlation. In fact, for a given angle ${\eta_{{\mathbf{a}}{\mathbf{b}}}}$ between ${\bf a}$ and ${\bf b}$, all four combinations, ${\mathscr{A}\mathscr{B}=+\,+}$, ${+\,-}$, ${-\,+}$, and ${-\,-}$, must occur in the results observed by Alice and Bob, with the probabilities given by
\begin{equation}
P^{+-}(\eta_{{\mathbf{a}}{\mathbf{b}}})=P\{{\mathscr A}=+1,\;{\mathscr B}=-1
\;|\;\eta_{{\mathbf{a}}{\mathbf{b}}}\}\;=\frac{1}{2}\cos^2\left(\frac{\eta_{{\mathbf{a}}{\mathbf{b}}}}{2}\right)
\end{equation}
and
\begin{equation}
P^{++}(\eta_{{\mathbf{a}}{\mathbf{b}}})=P\{{\mathscr A}=+1,\;{\mathscr B}=+1
\;|\;\eta_{{\mathbf{a}}{\mathbf{b}}}\}\,=\frac{1}{2}\sin^2\left(\frac{\eta_{{\mathbf{a}}{\mathbf{b}}}}{2}\right)\!,
\end{equation}
together with
\begin{equation}
P^{-+}(\eta_{{\mathbf{a}}{\mathbf{b}}})=P^{+-}(\eta_{{\mathbf{a}}{\mathbf{b}}}) \;\;\;\text{and}\;\;\;
P^{--}(\eta_{{\mathbf{a}}{\mathbf{b}}})=P^{++}(\eta_{{\mathbf{a}}{\mathbf{b}}}).
\end{equation}
Using these probabilities, the average or expected value of the product ${\mathscr{A}({\bf a},\,\lambda)\,\mathscr{B}({\bf b},\,\lambda)}$ can be easily worked out as:
\begin{align}
\!\!\!\!{\cal E}_{\rm Q.M.}({\bf a},\,{\bf b})\,&=\lim_{\,n\,\gg\,1}\left[\frac{1}{n}\sum_{k\,=\,1}^{n}\,{\mathscr A}({\bf a},\,\lambda^k)\;{\mathscr B}({\bf b},\,\lambda^k)\right] \notag \\
&=\,\frac{({\mathscr A}{\mathscr B}=+1)\times P^{++}\,+\,({\mathscr A}{\mathscr B}=+1)\times P^{--}\,+\,({\mathscr A}{\mathscr B}=-1)\times P^{+-}\,+\,({\mathscr A}{\mathscr B}=-1)\times P^{-+}}{P^{++}\,+\,P^{--}\,+\,P^{+-}\,+\,P^{-+}} \notag \\
&=\,\frac{P^{++}\,+\,P^{--}\,-\,P^{+-}\,-\,P^{-+}}{P^{++}\,+\,P^{--}\,+\,P^{+-}\,+\,P^{-+}} \notag \\
&=\,\frac{\frac{1}{2}\sin^2\left(\frac{\eta_{{\bf a}{\bf b}}}{2}\right)\,
+\,\frac{1}{2}\sin^2\left(\frac{\eta_{{\bf a}{\bf b}}}{2}\right)\,
-\,\frac{1}{2}\cos^2\left(\frac{\eta_{{\bf a}{\bf b}}}{2}\right)\,
-\,\frac{1}{2}\cos^2\left(\frac{\eta_{{\bf a}{\bf b}}}{2}\right)}{\frac{1}{2}\sin^2\left(\frac{\eta_{{\bf a}{\bf b}}}{2}\right)\,
+\,\frac{1}{2}\sin^2\left(\frac{\eta_{{\bf a}{\bf b}}}{2}\right)\,
+\,\frac{1}{2}\cos^2\left(\frac{\eta_{{\bf a}{\bf b}}}{2}\right)\,
+\,\frac{1}{2}\cos^2\left(\frac{\eta_{{\bf a}{\bf b}}}{2}\right)} \notag \\
&=\,\sin^2\left(\frac{\eta_{{\bf a}{\bf b}}}{2}\right)\,
-\,\cos^2\left(\frac{\eta_{{\bf a}{\bf b}}}{2}\right)\, \notag \\
&=\,-\,\cos\left(\eta_{{\bf a}{\bf b}}\right)\equiv-{\bf a}\cdot{\bf b}. \label{calcul}
\end{align}

These quantum mechanical predictions seem impossible to reconcile with the Dr.~Bertlmann's socks type correlation that respects Einstein's conceptions of locality, realism, and preexisting properties \cite{EPR}\cite{Howard-2019}. And yet, such a reconciliation is precisely the goal of this paper, and it is precisely what has been demonstrated in Refs.~\cite{Christian}, \cite{RSOS} and \cite{Christian2014}. The central idea presented in these publications is the following: As long as we insist on modeling the physical space as ${{\rm I\!R}^3}$ and insist on using ``vector algebra" as customarily employed in the Bell-test experiments \cite{Clauser}, the desired reconciliation is impossible \cite{disproof}. However, if we model the physical space as a quaternionic 3-sphere, ${S^3}$, and use Geometric Algebra~\cite{Clifford}, then reconciliation is not only possible but becomes inevitable, because the sign-flips such as (\ref{2}) occur naturally within\break ${S^3}$. They are induced by the M\"obius-like twists in the U(1) bundle over ${S^2}$ that constitute the quaternionic 3-sphere \cite{Eguchi}. As counterintuitive as this may seem, within ${S^3}$ it is possible to find a right-handed glove in one’s pocket and be certain that the glove forgotten at home must also be right-handed, at least for some of the measurement directions. Consequently, within $S^3$ the components of spins exist before they are measured, in line with Einstein's conception of local realism. To understand this affair better, in the next section we first demonstrate it using a fictitious toy model.

\section{Two-dimensional Analogue of the singlet Correlations}\label{II}

\begin{figure}
\hrule
\vspace{-0.85cm}
\scalebox{1.3}{
\begin{pspicture}(-2.0,-0.5)(2.0,5.2)

\epsfig{figure=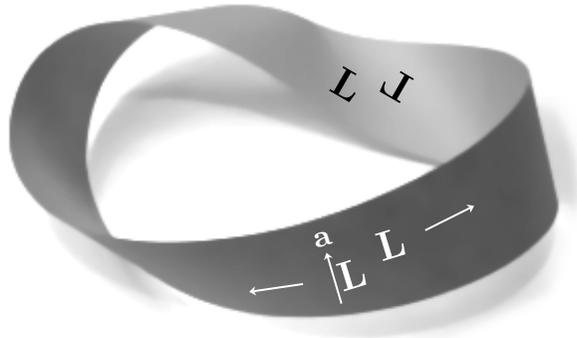,height=102pt,width=169.8pt}

\rput{-30}(-2.5,2.74){\large{\bf L}}

\rput{-30}(-2.0,2.74){\reflectbox{\large{\bf L}}}

\rput{17}(-2.05,1.11){\color{white}{\large{\bf L}}}

\rput{17}(-2.45,0.80){\color{white}{\large{\bf L}}}

\psline[linewidth=0.2mm,linecolor=white,arrowinset=0.3,arrowsize=1pt 3,arrowlength=0.7]{->}(-2.557,0.50)(-2.71,1.02)

\psline[linewidth=0.2mm,linecolor=white,arrowinset=0.3,arrowsize=1pt 3,arrowlength=0.7]{->}(-1.7,1.25)(-1.2,1.5)

\psline[linewidth=0.2mm,linecolor=white,arrowinset=0.3,arrowsize=1pt 3,arrowlength=0.7]{->}(-2.95,0.7)(-3.5,0.62)

\rput{17}(-2.75,1.16){\color{white}{\small ${\bf a}$}}

\end{pspicture}}
\vspace{-0.13cm}
\hrule
\caption{In the two-dimensional world of Alice and Bob two congruent shapes may become incongruent relative to each other.\break}
\vspace{0.23cm}
\label{fig-not-1}
\hrule
\end{figure}

Suppose Alice and Bob are two-dimensional creatures living in a two-dimensional, one-sided world resembling a M\"obius strip \cite{Walker}, entirely oblivious to the third dimension we take for granted (cf. Fig.~\ref{fig-not-1}). Suppose further that they discover certain correlations between the results of their observations that appear to be much stronger than any previously observed correlations, and their strength appears to be explainable only in non-local and/or non-realistic terms. However, being scientists, Alice and Bob strive to uncover a hypothesis that could explain the correlations in purely local and realistic terms. They may hypothesize, for example, that they are, in fact, living in a M\"obius world, embedded in a higher-dimensional space ${{\rm I\!R}^3}$ \cite{Walker}. Our goal in this section is to illustrate how such a hypothesis would explain their observed correlations in purely local-realistic terms. In the later sections we will relate it to the hypothesis we have advanced in Refs.~\cite{Christian}, \cite{RSOS} and \cite{Christian2014} to explain the strong correlations we observe in our three-dimensional world.

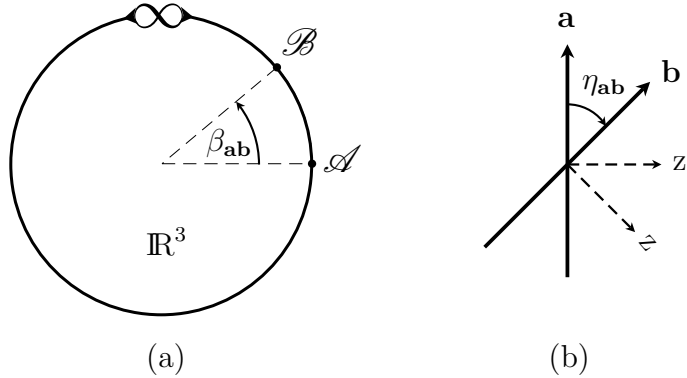
\begin{figure}
\hrule
\vspace{-0.25cm}
\scalebox{1}{
\begin{pspicture}(-3.4,-0.5)(8.5,5.2)

\put(-0.3,3.92){\huge${\infty}$}

\psline[linewidth=0.2mm,linecolor=white,arrowinset=0.3,arrowsize=1pt 3,arrowlength=0.7]{->}(-8.557,0.50)(-8.71,1.02)

\rput{17}(-8.75,1.16){\color{white}{\small ${\bf a}$}}

\psarc[linewidth=0.4mm,arrowinset=0.3,arrowsize=2pt 4,arrowlength=0.8]{>-<}(0.1,2.1){2.0}{99.12}{82.98}

\put(0.7,2.28){\large${\beta_{\bf ab}}$}

\psarc[showpoints=true,linewidth=0.2mm,linestyle=none]{*-*}(0.11,2.11){1.995}{0}{40}

\psarc[arrowinset=0.3,arrowsize=1pt 3,arrowlength=1]{->}(0.1,2.1){1.3}{0}{40}

\put(2.25,2.0){\Large ${\mathscr A}$}

\put(1.7,3.55){\Large ${\mathscr B}$}

\put(-0.1,0.9){\large ${{\rm I\!R}^3}$}

\put(-0.1,-0.6){\large (a)}

\psline[linewidth=0.5mm,arrowinset=0.3,arrowsize=1pt 3,arrowlength=1]{<-}(5.5,3.7)(5.5,0.6)

\psline[linewidth=0.5mm,arrowinset=0.3,arrowsize=1pt 3,arrowlength=1]{<-}(6.6,3.2)(4.4,1.0)

\psline[linewidth=0.3mm,linestyle=dashed,arrowinset=0.3,arrowsize=1pt 3,arrowlength=1]{->}(5.5,2.1)(6.75,2.1)

\psline[linewidth=0.3mm,linestyle=dashed,arrowinset=0.3,arrowsize=1pt 3,arrowlength=1]{->}(5.5,2.1)(6.4,1.2)

\put(6.9,2.02){\large ${\rm z}$}

\rput{-50}(6.57,1.06){\large ${\rm z}$}

\put(5.37,3.93){\large ${\bf a}$}

\put(6.75,3.2){\large${\bf b}$}

\psarc[arrowinset=0.3,arrowsize=1pt 3,arrowlength=1]{<-}(5.5,2.3){0.6}{31.2}{90}

\put(5.7,3.1){\large${\eta_{\bf ab}}$}

\put(5.25,-0.6){\large (b)}

\end{pspicture}}
\vspace{0.7cm}
\hrule
\caption{(a) The aerial view of the M\"obius world of Alice and Bob. The distance between their observation posts is given by the angle ${0\leq\beta_{\bf ab}\leq 2\pi}$. (b) The cross-sectional view of the world. The twist in the strip is characterized by
the angle ${0\leq\eta_{\bf ab}\leq\pi}$.}
\label{fig-not-2}
\vspace{0.2cm}
\hrule
\end{figure}

To this end, let Alice and Bob choose directions ${\bf a}$ and ${\bf b}$ to perform two independent sets of experiments, conducted at remote locations from each other. Within their two-dimensional world the vectors ${\bf a}$ and ${\bf b}$ could only have two coordinate components, but Alice and Bob hypothesize that perhaps their vectors also have third components, pointing ``outside'' of their own one-sided world. Let the twisting angle between these external components be denoted by ${\eta_{\bf ab}}$, as shown in Fig.~\ref{fig-not-2}~(b) below. Now the experiments Alice and Bob have been performing are exceedingly simple. It so happens that within their two-dimensional world wherever they set up their posts and choose directions for their measurements, they start receiving a stream of L-shaped patterns. Upon receiving each such pattern they record whether it has a left-handed L-shape or a right-handed L-shape. They determine this by aligning the longer arm of each pattern along their chosen measurement direction, with the shorter arm hanging in the opposite direction, as shown in Fig.~\ref{fig-not-1}. It is then easy for them to see whether the pattern has a left-handed L-shape or a right-handed L-shape. If it turns out to have a left-handed L-shape, Alice and Bob record the number ${-1}$ in their logbooks, and if it turns out to have a right-handed L-shape, they record the number ${+1}$ in their logbooks. What they always find in any such experiment involving a large number of patterns is that the sum total of all the numbers they end up recording, independently of each other, always adds up to zero. In other words, the L-shaped patterns they both
ceaselessly receive are always evenly distributed between left-handed patterns and right-handed patterns. But when they get together at the end of the day and compare the entries in their logbooks, they find that their observations are strongly correlated. They find, in fact, that the correlations among their observations can be expressed in terms of the angle ${\eta_{\bf ab}}$ shown in Fig.~\ref{fig-not-2}~(b), as
\begin{equation}
{\cal E}({\bf a},\,{\bf b})\,=\,-\cos\eta_{\bf ab}\,.\label{exeta}
\end{equation}

To explain these correlations in local-realistic terms, Alice and Bob hypothesize that the three-dimensional space external to their own is occupied by a mischievous gremlin they cannot see, who is hurling complementary L-shaped patterns towards them in a steady stream. What is more, this gremlin has a habit of making a random but evenly distributed choice between hurling a pattern towards his right or his left, with a complementary pattern hurled in the opposite direction. Each choice by the gremlin thus constitutes an evenly distributed random  hidden variable ${\lambda=+\,1}$ or ${-1}$, determining both the initial states of the patterns as well as the measurement results of Alice and Bob:
\begin{align}
{\mathscr A}({\bf a},\,{\lambda})\,&:=\,
\begin{cases}
+\,1\;\;\;\;\;{\rm if} &\lambda\,=\,+\,1 \\
-\,1\;\;\;\;\;{\rm if} &\lambda\,=\,-\,1
\end{cases} \label{mob-1}
\end{align}
and
\begin{align}
\;\;\;\;\;\;\;{\mathscr B}({\bf b},\,{\lambda})\,&:=\,
\begin{cases}
-\,1\;\;\;\;\;{\rm if} &\lambda\,=\,+\,1 \\
+\,1\;\;\;\;\;{\rm if} &\lambda\,=\,-\,1\,.\;\;\;\;\;\;
\end{cases} \label{mob-2}
\end{align}
These results are thus local, realistic, and deterministically determined. In fact, they respect the following properties.
\begin{itemize}
\item \underbar{Locality}: Apart from the common cause ${\lambda}$, the result ${{\mathscr A}=\pm1}$ depends {\it only} on the measurement direction ${\bf a}$, chosen freely by Alice, regardless of Bob's actions. And, similarly, apart from the common cause ${\lambda}$, the result ${{\mathscr B}=\pm1}$ depends {\it only} on the measurement direction ${\bf b}$, chosen freely by Bob, regardless of Alice's actions. In particular, the function ${{\mathscr A}({\bf a},\,\lambda)}$ {\it does not} depend on either ${\bf b}$ or ${\mathscr B}$ and the function ${{\mathscr B}({\bf b},\,\lambda)}$ {\it does not} depend on either ${\bf a}$ or ${\mathscr A}$. Moreover, the common cause or hidden variable ${\lambda}$ does not depend on either ${\bf a}$, ${\bf b}$, ${\mathscr A}$, or ${\mathscr B}$. The hidden variable theory hypothesized by Alice and Bob is therefore local in the sense espoused by Einstein. 
\item \underbar{Realism}: The L-shaped patterns preexist as either right- or left-handed regardless of anyone observing them. Alice, for example, can predict with confidence what result Bob would obtain if he does make a measurement, by simply being aware of her own result of measurement {\it and} the geometrical structure of the world they live in. 
\item \underbar{Determinism}: Because the results ${{\mathscr A}({\bf a},\,\lambda)}$ and ${{\mathscr B}({\bf b},\,\lambda)}$ of measurements made by Alice and Bob are definite when the vectors ${\bf a}$ and ${\bf b}$ and variable ${\lambda}$ are specified, their hypothesized hidden variable theory is deterministic.
\item \underbar{Non-contextuality}: Since the numbers ${\mathscr A}$ and ${\mathscr B}$ are read off simply by aligning the received patterns along the freely chosen vectors ${\bf a}$ and ${\bf b}$ and noting whether the patterns are right- or left-handed, the values or directions of the vectors ${\bf a}$ and ${\bf b}$ are of no real significance for the measurements of Alice and Bob that are made at their respective locations on the M\"obius strip. Any local direction ${\bf a}$ chosen by Alice, for example, would give the same answer to the question whether a received pattern is right-handed or left-handed. Moreover, the z-components of ${\bf a}$ and ${\bf b}$ (inaccessible to the two-dimensional Alice and Bob) also have no significance as far as their local measurements are concerned. The hidden variable theory hypothesized by Alice and Bob is thus non-contextual.
\end{itemize}

Given the above properties and the definitions (\ref{mob-1}) and (\ref{mob-2}), it is clear that the correlation between the results obtained independently by Alice and Bob will be entirely analogous to that between Dr.~Bertlmann's socks. This may give the impression that the product ${{\mathscr A}{\mathscr B}}$ of their results will therefore always remain at the fixed value of ${-1}$, just as in Dr.~Bertlmann's socks type correlation. But that impression would be wrong. The product, in fact, will fluctuate inevitably between the values ${-1}$ and ${+1}$:
\begin{equation}
{\mathscr A}{\mathscr B} \,\in\, \{\,-1,\,+1\,\}.
\end{equation}
To appreciate this, recall that in the one-sided world of M\"obius strip two congruent left-handed figures may not always remain congruent (cf. Fig.~\ref{fig-not-1}) \cite{Walker}. If one of the two left-handed figures moves around the strip an odd number of times {\it relative} to the stay-at-home figure, then it becomes right-handed, and hence incongruent with the stay-at-home figure \cite{Walker}. And the same would be true for two right-handed figures. This is quite easy to verify by making a model of a M\"obius strip from a strip of paper. If one starts with two incongruent L-shaped cardboard cutouts and moves one of them around the strip relative to the other, then after an odd number of revolutions the two cutouts become congruent with one another. Consequently, the value of the corresponding product ${{\mathscr A}{\mathscr B}}$ representing congruence or incongruence of the two L-shaped figures would change from ${-1}$ at the start of the trip to ${+1}$ at the end of the trip.

The reason for this, of course, is that --- unlike in a cylindrical strip --- there is a twist in the M\"obius strip. And this twist is ultimately responsible for the strong correlations observed by Alice and Bob. To understand this, suppose the gremlin happens to hurl a right-handed pattern towards Alice and a left-handed pattern towards Bob ({\it i.e.}, suppose ${\lambda=+1}$). What are the chances that Alice would then record the number ${+1}$ in her logbook whereas Bob would record the number ${-1}$ in his logbook? The answer to this question would depend, in fact, on where Alice and Bob are situated within the M\"obius strip, which can be parameterized in terms of the external angle ${\beta_{\bf ab}}$ quantifying the cylinder that would result if we imagine that the twist in it has been removed, as shown in Fig.~\ref{fig-not-2}~(a). If the posts of Alice and Bob happen to be almost next to each other, then their patterns are unlikely to undergo relative handedness transformation, and then Alice and Bob would indeed record ${+1}$ and ${-1}$, respectively, yielding the product ${{\mathscr A}{\mathscr B}=-1}$. If, however, Bob's post is almost a full circle away from Alice's post, then both Alice and Bob would record ${+1}$ with near certainty, because then Bob's pattern would have transformed into a right-handed pattern relative to Alice's pattern with near certainty, yielding the product ${{\mathscr A}{\mathscr B}=+1}$. For all intermediate angles the probability of the two patterns having undergone relative handedness transformation would be equal to ${{\beta_{\bf ab}}/2\pi}$, and the probability of the same two patterns {\it not} having undergone relative handedness transformation would be equal to ${(2\pi - {\beta_{\bf ab}})/2\pi}$. Thus all four possible combinations of outcomes, ${+\,+}$, ${+\,-}$, ${-\,+}$, and ${-\,-}$, would be observed by Alice and Bob, just as in the derivation (\ref{calcul}) discussed above. The corresponding correlations among their results would therefore work out as
\begin{align}
{\cal E}({\bf a},\,{\bf b})&\,=\,\lim_{\,n\,\gg\,1}\left[\frac{1}{n}\sum_{k\,=\,1}^{n}\,{\mathscr A}({\bf a},\,{\lambda}^k)\,
{\mathscr B}({\bf b},\,{\lambda}^k)\right] \notag \\
&\,=\,\frac{({\mathscr A}{\mathscr B}=+1)\times\frac{\beta_{\bf ab}}{2\pi}
\,+\,({\mathscr A}{\mathscr B}=-1)\times\frac{(2\pi-\beta_{\bf ab})}{2\pi}}{\frac{\beta_{\bf ab}}{2\pi}+\frac{(2\pi-\beta_{\bf ab})}{2\pi}} \notag \\
&\,=\,\frac{(+1)\times\beta_{\bf ab}\,+\,(-1)\times(2\pi-\beta_{\bf ab})}{2\pi} \notag \\
&\,=\,-1\,+\,\frac{1}{\pi}\;\beta_{\bf ab}\;\;\;\,\left(i.e.,\;\,\text{linear within the cylinder}\right)\!.\label{mobcorfi}
\end{align}
The validity of this result is straightforward to check by substituting ${\beta_{\bf ab}=0}$,
${\pi}$, and ${2\pi}$ to obtain ${{\cal E}({\bf a},\,{\bf b})=-1}$, ${0}$, and ${+1}$, respectively. However, having both ${{\mathscr A}{\mathscr B}=-1}$ and ${{\mathscr A}{\mathscr B}=+1}$ is not sufficient to explain the correlations (\ref{exeta}).

Fortunately, the above correlations are expressed in terms of the external angle ${\beta_{\bf ab}}$, which Alice and Bob can measure in radians as a distance between their observation posts. Therefore, as a final step towards explaining the correlations (\ref{exeta}) they must work out the distance ${\beta_{\bf ab}}$ in terms of the angle ${\eta_{\bf ab}}$ {\it intrinsic} to their world, which is not a difficult task. Recalling the properties of the M\"obius strip and noticing from its cross-sectional view in Fig.~\ref{fig-not-2}~(b) that the angle ${\eta_{\bf ab}}$ characterizes the {\it intrinsic} twist in the strip, it is easy to see that the two angles are, in fact, related as
\begin{equation}
\beta_{\bf ab}\,=\,\pi\left\{1-\cos\eta_{\bf ab}\right\}.\label{relaangle}
\end{equation}
This relation is the {\it defining} ${\,}$relation of the M\"obius world of Alice and Bob. Substituting it into Eq.~(\ref{mobcorfi}) they therefore arrive at
\begin{align}
{\cal E}({\bf a},\,{\bf b})\,&=\,-1\,+\,\frac{1}{\pi}\;\beta_{\bf ab}\, \notag \\
&=\,-1\,+\,\frac{1}{\pi}\times\left[\,\pi\left\{1-\cos\eta_{\bf ab}\right\}\right] \notag \\
&=\,-\cos\eta_{\bf ab}\,,
\end{align}
thereby explaining the mystery of the strong correlations they observe in their two-dimensional, one-sided world. 

In summary, what provides an unmistakable signature verifying their hypothesis in the experiment performed by our two-dimensional Alice and Bob are the three averages, $\Bigl\langle\,\mathscr{A}({\bf a})\,\Bigr\rangle=0$, $\Bigl\langle\,\mathscr{B}({\bf b})\,\Bigr\rangle=0$, and $\Bigl\langle\,\mathscr{A}({\bf a})\,\mathscr{B}({\bf b})\,\Bigr\rangle = -\cos\eta_{\bf ab}$, of the individual measurement results $\mathscr{A}({\bf a})$ and $\mathscr{B}({\bf b})$ and the joint measurement results $\mathscr{A}({\bf a})\,\mathscr{B}({\bf b})$ observed by them. Now, as noted in Eq.~(\ref{mobcorfi}) and the paragraph that includes it, when the twist in the M\"obius world is ignored, then it reduces to a cylinder with a flat intrinsic geometry. But, in that case, the averages of the individual measurement results $\mathscr{A}({\bf a})$ and $\mathscr{B}({\bf b})$ as defined in equations (\ref{mob-1}) and (\ref{mob-2}) remain unchanged. In that case, Alice and Bob would still observe $\Bigl\langle\,\mathscr{A}({\bf a})\,\Bigr\rangle=0$ and $\Bigl\langle\,\mathscr{B}({\bf b})\,\Bigr\rangle=0$.
However, as noted in Eq.~(\ref{mobcorfi}), the average of the joint measurement results changes from the sinusoidal correlations $\Bigl\langle\,\mathscr{A}({\bf a})\,\mathscr{B}({\bf b})\,\Bigr\rangle = -\cos\eta_{\bf ab}$ to linear correlations $\Bigl\langle\,\mathscr{A}({\bf a})\,\mathscr{B}({\bf b})\,\Bigr\rangle = -1+\frac{1}{\pi}\,\beta_{\bf ab}$.

Needless to say, as instructive as it is, this fictitious analogue of the singlet correlations cannot be taken too seriously. As we shall see in the next section, it helps us understand the real world correlations to a certain extent, but there are analogies as well as disanalogies between the two worlds. For example, although the twists within the geometrical structures of both worlds are responsible for the sinusoidal correlations, unlike the M\"obius strip the 3-sphere, or ${S^3}$, is an {\it orientable} manifold \cite{Misner}\cite{Penrose}. Thus, it is not the non-orientability, but the consistency of orientation (or handedness) within the 3-sphere that brings about the variations ${+\,+}$, ${+\,-}$, ${-\,+}$, and ${-\,-}$ in the observed results of Alice and Bob. In other words, while the twist in the M\"obius strip responsible for the strong correlations is an {\it extrinsic} twist in the geometry of the strip, the twists in the Hopf bundle of ${S^3}$ are {\it intrinsic} to ${S^3}$ \cite{Christian}\cite{RSOS}\cite{Christian2014}. Consequently, unlike in the M\"obius world where relative handedness of two L-shapes depends on the revolutionary distance between them, in the real world the relative handedness of the quaternions that constitute the 3-sphere reflects their intrinsic spinorial characteristics, independently of any distance between them. In the next section we look at these facts more closely. 

\section{Singlet Correlations in the quaternionic world}\label{III}

\begin{figure*}[t]
\hrule
\vspace{0.2cm}
\scalebox{1.6}{
\begin{pspicture}(-0.7,-2.3)(-6.9,-4.95)

\begin{rotate}{-180}

\pscurve[linewidth=0.2mm,linestyle=dashed](3.25,2.64)(3.75,2.73)(5.05,3.6)(4.0,3.6)(2.85,2.75)(3.25,2.64)

\rput{180}(3.45,4.7){\scriptsize {${\bf r}$}}

\psline[linewidth=0.2mm,linestyle=dashed,arrowinset=0.3,arrowsize=2pt 3,arrowlength=2]{<-}(3.9,3.15)(3.5,4.5)

\pscurve[linewidth=0.2mm]{-}(3.65,3.25)(3.51,3.07)(4.3,3.25)(4.23,3.32)

\psline[linewidth=0.2mm,arrowinset=0.3,arrowsize=2pt 3,arrowlength=2]{->}(3.9,2.985)(4.0,3.005)

\rput{194}(4.3,2.65){\scriptsize {${I\cdot{\bf r}}$}}

\end{rotate}
\end{pspicture}}
\vspace{0.2cm}
\hrule
\caption{A unit bivector represents an equatorial point of a quaternionic 3-sphere. As shown in the figure, a bivector is a {\it directed} number, characterized by only three abstract properties: (1) a magnitude, (2) a direction (specified by a vector orthogonal to it), and (3) a sense of rotation --- {\it i.e.}, clockwise (${-}$) or counterclockwise (${+}$). Neither the depicted oval shape of its plane, nor its axis of rotation ${\bf r}$, is an intrinsic part of the bivector ${{\bf J}({\bf r}):=I\cdot{\bf r}}$. The bivector ${{\bf J}({\bf r})}$ thus specifies ${\pm 1}$ spin about ${\bf r}$.\break}
\label{fig-5}
\hrule
\end{figure*}
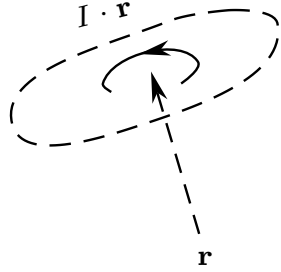

Learning from the two-dimensional Alice and Bob, we now hypothesize that {\it we live in a quaternionic 3-sphere}, or in ${S^3}$. This is by no means an {\it ad hoc} hypothesis, not the least because ${S^3}$ happens to be the spatial part of one of the well known cosmological solutions of Einstein's field equations of general relativity, corresponding to a closed Universe with a positive curvature \cite{Christian2014}. Indeed, the cosmic microwave background (CMB) spectra recently mapped by the space observatory {\it Planck} now prefers a positive curvature at more than 99\% confidence level \cite{closed}\cite{Handley}. Moreover, it is well known that representations of rotations using quaternions can most effectively capture the fact that the state of any rotating body in physical space depends in general not only on its local configuration but also on its topological relation to the rest of the Universe \cite{Misner}. While the former feature of the physical space is familiar from everyday life, the latter feature can also be demonstrated by a simple rope trick, or Dirac's belt trick \cite{Penrose}\cite{Hartung}. The appropriate operational question in this context is: Whether rotating bodies in physical space respect ${2\pi}$ periodicity or ${4\pi}$ periodicity? Consider, for example, a rock in an otherwise empty universe. If it is rotated by ${2\pi}$ radians about some axis, then there is no reason to doubt that it will return back to its original state with no discernible effects \cite{Hartung}. This, however, cannot be expected if there is at least one other object present in the universe \cite{Hartung}. The rock will then have to rotate by another ${2\pi}$ radians ({\it i.e.}, a total of ${4\pi}$ radians) to return back to its original state relative to that other object, as proved by the twist in the belt in Dirac's belt trick. The twist shows that what is an identity transformation for an isolated object is {\it not} an identity transformation for an object that is rotating relative to other objects \cite{Misner}\cite{Hartung}. We can quantify this\break loss of identity by adapting a spinor representation of rotations using a set of unit quaternions, known as a 3-sphere. Let the configuration space of all possible rotations of the rock be represented by the set ${S^3}$ of unit quaternions \cite{Christian}:
\begin{equation}
S^3:=\left\{\,{\bf q}(\psi,\,{\bf r}):=\exp\left[\,{\bf J}({\bf r})\,\frac{\psi}{2}\,\right]
\Bigg|\;||\,{\bf q}(\psi,\,{\bf r})\,||^2=1\right\}\!, \label{non}
\end{equation}
where ${{\bf J}({\bf r})}$ is a bivector (or a pure quaternion; cf. Fig.~\ref{fig-5}) rotating about ${{\bf r}\in{\rm I\!R}^3}$ with the rotation angle ${\psi}$ in the range ${0\leq\psi < 4\pi}$. Throughout this paper we will follow the notations, conventions, and terminology of Geometric Algebra \cite{Christian2014}\cite{Clifford}. Accordingly, ${{\bf J}({\bf r})\in S^2\subset S^3}$ can be parameterized by a unit vector ${{\bf r}=r_1\,{\bf e}_1+r_2\,{\bf e}_2+r_3\,{\bf e}_3\in{\rm I\!R}^3}$ as
\begin{align}
{\bf J}({\bf r})\,:=\,(\,I\cdot{\bf r}\,)\,&=\,r_1\,(\,I\cdot{\bf e}_1\,)
\,+\,r_2\,(\,I\cdot{\bf e}_2\,)\,+\,r_3\,(\,I\cdot{\bf e}_3\,) \notag \\
&=\,r_1\;{{\bf e}_2}\,\wedge\,{{\bf e}_3}
\,+\,r_2\;{{\bf e}_3}\,\wedge\,{{\bf e}_1}\,+\,r_3\;{{\bf e}_1}\,\wedge\,{{\bf e}_2}\,, \label{16}
\end{align}
with ${{\bf J}^2({\bf r})=-1}$. Here the trivector ${I:={\bf e}_1\wedge{\bf e}_2\wedge{\bf e}_3\equiv{\bf e}_1{\bf e}_2{\bf e}_3}$, with ${I^2=-1}$, represents a volume form in ${{\rm I\!R}^3}$, and ${\{{\bf e}_j}\,\wedge\,{{\bf e}_k\}}$ forms a bivector basis. Each configuration of the rock can be represented by a unit quaternion of the form
\begin{equation}
{\bf q}(\psi,\,{\bf r})\,=\,\cos\left(\frac{\psi}{2}\right)\,+\,{\bf J}({\bf r})\,\sin\left(\frac{\psi}{2}\right), \label{defi-2}
\end{equation}
with ${\psi}$ being its rotation angle from ${{\bf q}(0,\,{\bf r})=1}$ \cite{Hartung}\cite{Altmann}. Incidentally, quaternions are a left-handed set of bivectors \cite{Clifford}. More significantly for our purposes, it is easy to verify that ${{\bf q}(\psi,\,{\bf r})}$ respects the following rotational symmetries:
\begin{equation}
{\bf q}(\psi+4\kappa\pi,\,{\bf r})\,=\,+\,{\bf q}(\psi,\,{\bf r})\,\;\;\text{for}\;\,\kappa=0,1,2,3,\dots \label{18}
\end{equation}
and
\begin{equation}
{\bf q}(\psi+2\kappa\pi,\,{\bf r})\,=\,-\,{\bf q}(\psi,\,{\bf r})\,\;\;\text{for}\;\,\kappa=1,3,5,7,\dots \label{19}
\end{equation}
Thus ${{\bf q}(\psi,\,{\bf r})}$ correctly represents the state of a rock that returns to itself only after even multiples of a ${2\pi}$ rotation.

Now, given two unit vectors ${\bf x}$ and ${\bf y}$ and a rotation axis ${\bf r}$, each element of ${S^3}$ can be factorized into a product of corresponding bivectors ${{\bf J}({\bf x})}$ and ${{\bf J}({\bf y})}$ [which can be expanded in terms of basis bivectors as in Eq.~(\ref{16})], as follows:
\begin{equation}
{\bf q}(\eta_{{\bf x}{\bf y}},\,{\bf r})\,=\,-\,{\bf J}({\bf x})\,{\bf J}({\bf y})\,=\,-\,(I\cdot{\bf x})\,(I\cdot{\bf y})\,=\,{\bf x}\,{\bf y}\,=\,{\bf x}\cdot{\bf y}\,+\,{\bf x}\wedge{\bf y} \,=\,\cos(\,\eta_{{\bf x}{\bf y}})\,+\,{\bf J}({\bf r})\,\sin(\,\eta_{{\bf x}{\bf y}})\,, \label{20}
\end{equation}
where ${\eta_{{\bf x}{\bf y}}}$ is the angle between ${\bf x}$ and ${\bf y}$, ${{\bf x}\,{\bf y}}$ is the geometric product between ${\bf x}$ and ${\bf y}$, ${{\bf x}\wedge{\bf y}}$ is the wedge product between ${\bf x}$ and ${\bf y}$, and ${{\bf J}({\bf r})}$ is identified with ${\frac{{\bf x}\wedge{\bf y}}{||{\bf x}\wedge{\bf y}||}}$. Comparing Eqs.~(\ref{defi-2}) and (\ref{20}), we can now see that the rotation angle ${\psi}$ of the quaternion is twice the angle ${\eta_{{\bf x}{\bf y}}}$ between the vectors ${\bf x}$ and ${\bf y}$ in any factorization such as in Eq.~(\ref{20}):
\begin{equation}
\psi\,=\,2\,\eta_{{\bf x}{\bf y}}.
\end{equation}
Consequently, the fundamental spinorial sign changes \cite{Christian}\cite{Hartung} expressed in Eqs.~(\ref{18}) and (\ref{19}) can be expressed also as
\begin{equation}
{\bf q}(\eta_{{\bf x}{\bf y}}+2\kappa\pi,\,{\bf r})\,=\,+\,{\bf q}(\eta_{{\bf x}{\bf y}},\,{\bf r})\,\;\;\text{for}\;\,\kappa=0,1,2,3,\dots
\end{equation}
and
\begin{equation}
{\bf q}(\eta_{{\bf x}{\bf y}}+\kappa\pi,\,{\bf r})\,=\,-\,{\bf q}(\eta_{{\bf x}{\bf y}},\,{\bf r})\,\;\;\text{for}\;\,\kappa=1,3,5,7,\dots \label{23}
\end{equation}
This last equation, Eq.~(\ref{23}), exhibits a key relation that reduces the singlet correlations we observe in the real world to Dr.~Bertlmann's socks type classical correlations, because, as we shall soon see, it transforms ${{\mathscr A}{\mathscr B}=-1}$ to ${{\mathscr A}{\mathscr B}=+1}$.

To appreciate this, let us represent the spin angular momenta in a typical EPR-Bohm type experiment shown in Fig.~\ref{fig-1} by a set of unit bivectors \cite{Christian}\cite{RSOS}\cite{Christian2014}. They can be expressed in terms of graded bivector basis using the sub-algebra
\begin{equation}
L_{i}(\lambda)\,L_{j}(\lambda) \,=\,-\,\delta_{ij}\,-\,\sum_{k}\,\epsilon_{ijk}\,L_{k}(\lambda)\,, \label{wh-o8899}
\end{equation}
which span a tangent space at each point of ${S^3}$, with a choice of handedness ${\lambda=\pm\,1}$. Contracting this equation on both sides with the components ${a^{i}}$ and ${b^{j}}$ of arbitrary unit vectors ${\bf a}$ and ${\bf b}$ then gives the convenient bivector identity
\begin{equation}
{\bf L}({\bf a},\,\lambda)\,{\bf L}({\bf b},\,\lambda)\,=\,-\,{\bf a}\cdot{\bf b}\,-\,{\bf L}({\bf a}\times{\bf b},\,\lambda)\,, \label{50}
\end{equation}
where ${\mathbf{L}(\mathbf{a},\,\lambda):=a^iL_i(\lambda)}$ and ${\mathbf{L}(\mathbf{b},\,\lambda):=b^jL_j(\lambda)}$ are unit bivectors. The identity (\ref{50}) is the Pauli identity. It simply expresses a geometric product between the unit bivectors ${\mathbf{L}(\mathbf{a},\,\lambda)}$ and ${\mathbf{L}(\mathbf{b},\,\lambda)}$, representing the spin angular momenta
\begin{equation}
{\bf L}({\bf a},\,{\lambda})\,=\,\lambda\,I\,{\bf a}\,=\,\lambda\,I\cdot{\bf a}\,\equiv\,\lambda({\bf e}_1\wedge{\bf e}_2\wedge{\bf e}_3)\cdot{\bf a}\,=\,(\pm 1 \;\,\text{spin about the direction}\;\,{\bf a}) \label{12}
\end{equation}
and
\begin{equation}
{\bf L}({\bf b},\,{\lambda})\,=\,\lambda\,I\,{\bf b}\,=\,\lambda\,I\cdot{\bf b}\,\equiv\,\lambda({\bf e}_1\wedge{\bf e}_2\wedge{\bf e}_3)\cdot{\bf b} \,=\,(\pm 1 \;\,\text{spin about the direction}\;\, {\bf b}),\!\!\! \label{13}
\end{equation}
where the trivector property ${I^2=-1}$ and the unity of the vectors ${\bf a}$ and ${\bf b}$ ensure that ${{\bf L}^2({\bf a},\,\lambda)=-1={\bf L}^2({\bf b},\,\lambda)}$.

We are now in a position to derive the singlet correlations observed in the EPR-Bohm type experiments in elegant manner \cite{Christian2014}. To this end, following Bell's formalism \cite{Bell-1964}, consider measurement functions for Alice and Bob of the form
\begin{equation}
\pm\,1\,=\,{\mathscr A}({\bf a},\,\lambda^k)\!:
{\mathrm{I\!R}}^3\!\times\left\{\,\lambda^k\right\}\longrightarrow S^3\hookrightarrow{\mathrm{I\!R}}^4
\end{equation}
and
\begin{equation}
\pm\,1\,=\,{\mathscr B}({\bf b},\,\lambda^k)\!:
{\mathrm{I\!R}}^3\!\times\left\{\,\lambda^k\right\}\longrightarrow S^3\hookrightarrow{\mathrm{I\!R}}^4,\end{equation}
where the handedness ${\lambda^k=+1}$ or ${-1}$ for each run ${k}$ of the experiment considered in Fig.~\ref{fig-1}. More explicitly, let the bivectors ${-\,{\bf L}({\bf s}_1,\,\lambda^k)}$ and ${+\,{\bf L}({\bf s}_2,\,\lambda^k)}$ representing the spins emerging from a common source be detected by space-like separated detector bivectors ${{\bf D}({\bf a})=I\cdot{\bf a}}$ and ${{\bf D}({\bf b})=I\cdot{\bf b}}$, freely chosen by Alice and Bob, and defining their measurement results 
\begin{equation}
S^3\ni\,{\mathscr A}({\bf a},\,{\lambda^k})\,:=\lim_{{\bf s}_1\,\rightarrow\,{\bf a}}\left\{\,+\,{\bf q}(\eta_{{\bf a}{\bf s}_1},\,{\bf r}_1)\right\}\,\equiv\lim_{{\bf s}_1\,\rightarrow\,{\bf a}}\left\{-\,{\bf D}({\bf a})\,{\bf L}({\bf s}_1,\,\lambda^k)\right\} \,\xrightarrow[{\bf s}_1\,\to\,{\bf a}]\,\begin{cases}
+\,1\;\;\;\;\;{\rm if} &\lambda^k\,=\,+\,1 \\
-\,1\;\;\;\;\;{\rm if} &\lambda^k\,=\,-\,1
\end{cases} \Bigg\}\label{53}
\end{equation}
and
\begin{equation}
S^3\ni\,{\mathscr B}({\bf b},\,{\lambda^k})\,:=\lim_{{\bf s}_2\,\rightarrow\,{\bf b}}\left\{\,-\,{\bf q}(\eta_{{\bf s}_2{\bf b}},\,{\bf r}_2)\right\}\,\equiv\lim_{{\bf s}_2\,\rightarrow\,{\bf b}}\left\{+\,{\bf L}({\bf s}_2,\,\lambda^k)\,{\bf D}({\bf b})\right\} \,\xrightarrow[{\bf s}_2\,\to\,{\bf b}]\,\begin{cases}
-\,1\;\;\;\;\;{\rm if} &\lambda^k\,=\,+\,1 \\
+\,1\;\;\;\;\;{\rm if} &\lambda^k\,=\,-\,1
\end{cases} \Bigg\}, \label{54}
\end{equation}
where we have assumed the handedness ${\lambda^k}$ of ${S^3}$ to be a fair coin with 50/50 chance of being ${+1}$ or ${-\,1}$ at the moment of pair-creation, making the spinning bivector ${{\bf L}({\bf n},\,\lambda^k)}$ a random variable {\it relative} to any detector bivector such as ${{\bf D}({\bf n})=I\cdot{\bf n}}$,
\begin{equation}
{\bf L}({\bf n},\,\lambda^k)\,=\,\lambda^k\,{\bf D}({\bf n})\,\,\Longleftrightarrow\,\,{\bf D}({\bf n})\,=\,\lambda^k\,{\bf L}({\bf n},\,\lambda^k)\,. \label{55}
\end{equation}
The fact that orientation ${\lambda^k}$ of ${S^3}$ is a fair coin ensures that the results observed by Alice and Bob vanish on average:
\begin{equation}
\Bigl\langle\,{\mathscr A}({\bf a},\,\lambda^k)\,\Bigr\rangle\,=\,0\;\;\;\text{and}\;\;\; \Bigl\langle\,{\mathscr B}({\bf b},\,\lambda^k)\,\Bigr\rangle\,=\,0.
\end{equation}
It is also evident from the measurement functions ${{\mathscr A}({\bf a},\,{\lambda^k})}$ and ${{\mathscr B}({\bf b},\,{\lambda^k})}$ that their values are limiting scalar points, ${\pm1}$, of two of the quaternions, ${{\bf q}(\eta_{{\bf a}{\bf s}_1},\,{\bf r}_1):=-{\bf D}({\bf a}){\bf L}({\bf s}_1,\,\lambda^k)}$ and ${-\,{\bf q}(\eta_{{\bf s}_2{\bf b}},\,{\bf r}_2):=+\,{\bf L}({\bf s}_2,\,\lambda^k)\,{\bf D}({\bf b})}$. Consequently, they respect the geometry and topology of ${S^3}$ rather than those of ${{\mathrm{I\!R}^3}}$. Physically, the geometry of ${S^3}$ stems from the rotations of the two spin bivectors, ${-{\bf L}({\bf s}_1,\,\lambda^k)}$ and ${+{\bf L}({\bf s}_2,\,\lambda^k)}$, {\it relative} to the detector bivectors ${{\bf D}({\bf a})}$ and ${{\bf D}({\bf b})}$.

The important question now is: What is the value of the product ${{\mathscr A}{\mathscr B}\,}$? We can work out the value of ${{\mathscr A}{\mathscr B}}$ from the definitions (\ref{53}) and (\ref{54}) of ${{\mathscr A}({\bf a},\,{\lambda^k})}$ and ${{\mathscr B}({\bf b},\,{\lambda^k})}$ and the ``product of limits equal to limits of product'' rule:
\begin{align}
S^3\ni\,{\mathscr A}{\mathscr B}({\bf a},\,{\bf b},\,{\lambda^k})\,&=\,{\mathscr A}({\bf a},\,{\lambda^k})\,{\mathscr B}({\bf b},\,{\lambda^k}) \\
&=\,\left[\lim_{{\bf s}_1\,\rightarrow\,{\bf a}}\left\{\,+\,{\bf q}(\eta_{{\bf a}{\bf s}_1},\,{\bf r}_1)\right\}\right]\left[\lim_{{\bf s}_2\,\rightarrow\,{\bf b}}\left\{\,-\,{\bf q}(\eta_{{\bf s}_2{\bf b}},\,{\bf r}_2)\right\}\right] \\
&=\,\lim_{\substack{{\bf s}_1\,\rightarrow\,{\bf a} \\ {\bf s}_2\,\rightarrow\,{\bf b}}}\left\{\,-\,{\bf q}(\eta_{{\bf a}{\bf s}_1},\,{\bf r}_1)\,{\bf q}(\eta_{{\bf s}_2{\bf b}},\,{\bf r}_2)\right\} \\
&=\,\lim_{\substack{{\bf s}_1\,\rightarrow\,{\bf a} \\ {\bf s}_2\,\rightarrow\,{\bf b}}}\left\{\,-\,{\bf q}(\eta_{{\bf u}{\bf v}},\,{\bf r}_{0})\right\} \label{35}\\
&=\,-1, \label{36}
\end{align}
where
\vspace{-0.4cm}
\begin{equation}
\eta_{{\bf u}{\bf v}}\,:=\,\cos^{-1}\left\{({\bf a}\cdot{\bf s}_1)({\bf s}_2\cdot{\bf b})-({\bf a}\cdot{\bf s}_2)({\bf s}_1\cdot{\bf b})+({\bf a}\cdot{\bf b})({\bf s}_1\cdot{\bf s}_2)\right\} \label{38}
\end{equation}
\vspace{-0.2cm}
and
\begin{equation}
{\bf r}_{0}\,:=\,\frac{({\bf a}\cdot{\bf s}_1)({\bf s}_2\times{\bf b})\,+\,({\bf s}_2\cdot{\bf b})({\bf a}\times{\bf s}_1)\,-\,({\bf a}\times{\bf s}_1)\times({\bf s}_2\times{\bf b})}{||\,({\bf a}\cdot{\bf s}_1)({\bf s}_2\times{\bf b})\,+\,({\bf s}_2\cdot{\bf b})({\bf a}\times{\bf s}_1)\,-\,({\bf a}\times{\bf s}_1)\times({\bf s}_2\times{\bf b})\,||}
\;\;\xrightarrow[\;\;\,{\substack{{\bf s}_1\,\rightarrow\,{\bf a} \\ {\bf s}_2\,\rightarrow\,{\bf b}}}\;\;]{}\; \vec{\bf \,0}. \label{null}
\end{equation}
That the product of the quaternions ${{\bf q}(\eta_{{\bf a}{\bf s}_1},\,{\bf r}_1)}$ and ${{\bf q}(\eta_{{\bf s}_2{\bf b}},\,{\bf r}_2)}$ is yet another quaternion ${{\bf q}(\eta_{{\bf u}{\bf v}},\,{\bf r}_{0})}$ is not surprising, because the set ${S^3}$ defined in Eq.~(\ref{non}) is known to remain closed under multiplication \cite{Christian}\cite{Christian2014}\cite{disproof}. A product of any number\break of quaternions will result in yet another quaternion belonging to ${S^3}$. More importantly for our hypothesis, the product ${{\mathscr A}{\mathscr B}({\bf a},\,{\bf b},\,{\lambda^k})}$ is again a limiting scalar point, ${-1}$ in this case, of the quaternion ${-\,{\bf q}(\eta_{{\bf u}{\bf v}},\,{\bf r}_{0})}$ that also belongs to ${S^3}$.

The result (\ref{36}), namely ${{\mathscr A}{\mathscr B}=-1}$, suggests that if Alice finds spin to be ``up" at her station, then Bob is guaranteed to find spin to be ``down" at his station, precisely mimicking the perfect anti-correlation observed in Dr.~Bertlmann's socks type correlations. As we discussed in Section \ref{II}, this may give the impression that the product ${{\mathscr A}{\mathscr B}}$ of the results\break observed by Alice and Bob will always remain at the fixed value of ${-1}$. But let us not forget the relation (\ref{23}), namely
\begin{equation}
{\bf q}(\eta_{{\bf x}{\bf y}}+\kappa\pi,\,{\bf r})\,=\,-\,{\bf q}(\eta_{{\bf x}{\bf y}},\,{\bf r})\,\;\;\text{for}\;\,\kappa=1,3,5,7,\dots, \tag{\ref{23}}
\end{equation}
which any quaternion in ${S^3}$ --- including the quaternion ${-\,{\bf q}(\eta_{{\bf u}{\bf v}},\,{\bf r}_{0})}$ appearing in Eq.~(\ref{35}) as well as those appearing in the definitions (\ref{53}) and (\ref{54}) of the individual measurement results  ${\mathscr A}$ and ${\mathscr B}$ --- must respect, thereby altering the value of the product ${{\mathscr A}{\mathscr B}}$ \cite{Hartung}. Although Alice and Bob do not have access to the spin directions ${{\bf s}_1}$ and ${{\bf s}_2}$, they are free to choose (and change at will) the detector directions ${\bf a}$ and ${\bf b}$ appearing in Eq.~(\ref{38}) that defines the angle ${\eta_{{\bf u}{\bf v}}}$. Consequently, variations in the detector directions ${\bf a}$ and ${\bf b}$ will induce variations in the angle ${\eta_{{\bf u}{\bf v}}}$, which can be expressed as ${\eta_{{\bf u}{\bf v}}\rightarrow\eta_{{\bf u}{\bf v}}+\delta}$. For variation ${\delta=\kappa\pi}$, the quaternion ${-\,{\bf q}(\eta_{{\bf u}{\bf v}},\,{\bf r}_{0})}$ appearing in Eq.~(\ref{35}) will then change its sign from ${-\,{\bf q}(\eta_{{\bf u}{\bf v}},\,{\bf r}_{0})}$ to ${+\,{\bf q}(\eta_{{\bf u}{\bf v}},\,{\bf r}_{0})}$ for odd ${\kappa}$. As a result, the value of the product ${{\mathscr A}{\mathscr B}}$ will change from ${-1}$ to ${+1}$ for odd ${\kappa}$. We would thus have our cake ({\it i.e.}, Dr.~Bertlmann's socks type local-realistic interpretation of the correlations) and eat it too ({\it i.e.}, have the value of the product ${{\mathscr A}{\mathscr B}}$ fluctuate between ${-1}$ and ${+1}$). In other words, all four possible combinations of outcomes, ${+\,+}$, ${+\,-}$, ${-\,+}$, and ${-\,-}$, will be observed by Alice and Bob, just as in the derivation (\ref{calcul}) of the quantum correlations, despite the correlations being Dr.~Bertlmann's socks type correlations. Note, however, that $\kappa$ is not a hidden variable. It is a part of the properties of a physical space Alice and Bob live in.

These results are thus local, realistic, and deterministically determined. In fact, they respect the following properties.
\begin{itemize}
\item \underbar{Locality}: Apart from the common cause ${\lambda}$, the result ${{\mathscr A}=\pm1}$ depends {\it only} on the measurement direction ${\bf a}$, chosen freely by Alice, regardless of Bob's actions. And, similarly, apart from the common cause ${\lambda}$, the result ${{\mathscr B}=\pm1}$ depends {\it only} on the measurement direction ${\bf b}$, chosen freely by Bob, regardless of Alice's actions. In particular, the function ${{\mathscr A}({\bf a},\,\lambda)}$ {\it does not} depend on either ${\bf b}$ or ${\mathscr B}$ and the function ${{\mathscr B}({\bf b},\,\lambda)}$ {\it does not} depend on either ${\bf a}$ or ${\mathscr A}$. Moreover, the common cause or hidden variable ${\lambda}$ does not depend on either ${\bf a}$, ${\bf b}$, ${\mathscr A}$, or ${\mathscr B}$. The hypothesized hidden variable theory is thus local in the sense espoused by Einstein and formalized by Bell. 
\item \underbar{Realism}: Because the results 
${{\mathscr A}({\bf a},\,\lambda)}$ and ${{\mathscr B}({\bf b},\,\lambda)}$ defined in (\ref{53}) and (\ref{54}) preexist within ${S^3}$ as limiting scalar points of ${S^3}$, the values of all spin components also preexist within ${S^3}$ regardless of anyone measuring them. Alice,\break for example, can predict with certainty what result Bob would obtain {\it if} he happens to make a measurement, by simply being aware of her own result {\it and} the algebraic, geometrical, and topological properties of the ${S^3}$ world. 
\item \underbar{Determinism}: Because the results ${{\mathscr A}({\bf a},\,\lambda)}$ and ${{\mathscr B}({\bf b},\,\lambda)}$ of the measurements made by Alice and Bob are definite when vectors ${\bf a}$ and ${\bf b}$ and initial state ${\lambda}$ are specified, the hypothesized hidden variable theory is deterministic.
\item \underbar{Contextuality}: As depicted in Fig.~\ref{fig-1}, two different vectors ${\bf a}$ and ${\bf a'}$ chosen by Alice may correspond to the same measurable quantity ${\mathscr A}$, measured along different contexts, without reference to what is measured on particle 2; and likewise for Bob. Therefore the hypothesized hidden variable theory is locally (but not remotely) contextual.
\end{itemize}

\begin{figure*}[t]
\hrule
\scalebox{0.7}{
\begin{pspicture}(0.3,-3.9)(4.5,3.1)

\pscircle[linewidth=0.3mm,linestyle=dashed](-1.8,-0.45){2.6}

\psellipse[linewidth=0.3mm](-0.8,-0.45)(0.7,1.4)

\psellipse[linewidth=0.3mm,border=3pt](-2.4,-0.45)(1.4,0.4)

\pscurve[linewidth=0.3mm,border=3pt](-1.485,-0.35)(-1.48,-0.25)(-1.45,0.0)

\pscircle[linewidth=0.3mm](7.0,-0.45){1.7}

\psellipse[linewidth=0.2mm,linestyle=dashed](7.0,-0.45)(1.68,0.4)

\put(-4.4,1.27){{\Large ${S^3}$}}

\put(-2.0,1.2){{\Large ${h^{-1}(q)}$}}

\put(-3.7,-1.4){{\Large ${h^{-1}(p)}$}}

\put(7.43,0.67){{\Large ${q}$}}

\psdot*(7.2,0.79)

\put(6.3,0.43){{\Large ${p}$}}

\psdot*(6.1,0.43)

\put(8.5,-1.8){{\Large ${S^2}$}}

\put(6.1,-2.7){\large base space}

\put(1.9,0.7){\Large ${h:S^3\rightarrow S^2}$}

\pscurve[linewidth=0.3mm,arrowinset=0.3,arrowsize=3pt 3,arrowlength=2]{->}(1.2,0.25)(2.47,0.45)(3.74,0.45)(4.9,0.25)

\put(1.73,-0.45){\large Hopf fibration}

\pscurve[linewidth=0.3mm,arrowinset=0.3,arrowsize=3pt 3,arrowlength=2]{->}(4.9,-0.95)(3.74,-1.15)(2.47,-1.15)(1.2,-0.95)

\put(1.75,-1.8){\Large ${h^{-1}:S^2\rightarrow S^3}$}

\end{pspicture}}
\hrule
\caption{The tangled web of linked Hopf circles depicting the geometrical and topological non-trivialities of ${S^3}$. Locally, ${S^3}$ is a product space: ${S^2\times S^1}$. It is thus ${S^2}$ worth of circles. But each circle (or fiber) ${S^1}$ threads through every other circle in the\break bundle ${S^3}$ without sharing a single point with it, and projects down to a point such as $p$ on ${S^2}$ via the Hopf map ${h: S^3\rightarrow S^2}$.}
\vspace{0.25cm}
\label{fig4}
\hrule
\end{figure*}
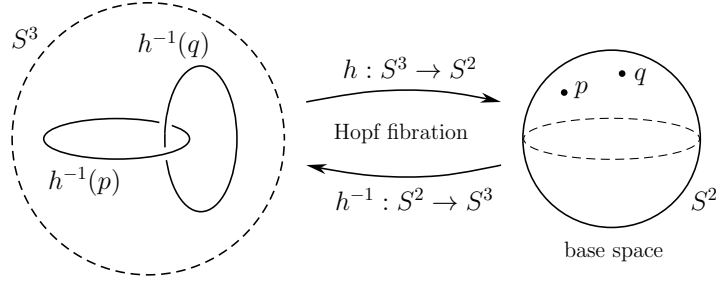

It is important to note, however, that having the value of the product ${{\mathscr A}{\mathscr B}}$ fluctuate between ${-1}$ and ${+1}$ does not, by itself, guarantee the strong correlations between the results ${\mathscr A}$ and ${\mathscr B}$. Indeed, calculations within ${{\mathrm{I\!R}^3}}$ \cite{Peres} are well known to predict correlations of the following linear form that also exhibit fluctuations between ${-1}$ and ${+1}$ in ${{\mathscr A}{\mathscr B}}$:
\begin{align}
{\cal E}({\bf a},\,{\bf b})\,=\,\lim_{\,n\,\gg\,1}\left[\frac{1}{n}\sum_{k\,=\,1}^{n}\,{\mathscr A}({\bf a},\,{\lambda}^k)\,
{\mathscr B}({\bf b},\,{\lambda}^k)\right]&=
\begin{cases}
-\,1\,+\,\frac{2}{\pi}\,\eta_{{\bf a}{\bf b}}
\;\;\;{\rm if} & \! 0 \leq \eta_{{\bf a}{\bf b}} \leq \pi \\
\\
+\,3\,-\,\frac{2}{\pi}\,\eta_{{\bf a}{\bf b}}
\;\;\;{\rm if} & \! \pi \leq \eta_{{\bf a}{\bf b}} \leq 2\pi\,, 
\end{cases}  \label{dismee}
\end{align}
where ${\eta_{{\bf a}{\bf b}}}$ is half the rotation angle ${\psi}$ between the directions ${\bf a}$ and ${\bf b}$ (for a plot of the above correlations see Fig.~\ref{fig6}). 

The difference between the above correlations and those derived in Eq.~(\ref{calcul}) is a consequence of the global algebraic, geometrical, and topological properties of ${S^3}$ \cite{Christian}\cite{RSOS}\cite{Christian2014}. While the individual quaternions ${{\bf q}(\eta_{{\bf x}{\bf y}},\,{\bf r})}$ contribute to this difference by inducing the spinorial sign changes described in Eq.~(\ref{23}), it is the global algebraic, geometrical, and topological properties of the physical space they {\it collectively} constitute --- namely, of the space of all unit quaternions
\begin{equation}
S^3:=\left\{\,{\bf q}(\eta_{{\bf x}{\bf y}},\,{\bf r}):=\cos(\,\eta_{{\bf x}{\bf y}})+{\bf J}({\bf r})\,\sin(\,\eta_{{\bf x}{\bf y}})\,
\Bigg|\;||\,{\bf q}(\eta_{{\bf x}{\bf y}},\,{\bf r})\,||^2=1\right\}, \label{three}
\end{equation}
that are responsible for producing the sinusoidal rather than linear correlations. Locally, in the topological sense, ${S^3}$ is a product space: ${S^2\times S^1}$ \cite{Ryder}. This is easy to see from the above definition. The axis of rotation ${{\bf r}}$ of the quaternion ${{\bf q}(\eta_{{\bf x}{\bf y}},\,{\bf r})}$ is a unit vector in ${{\rm I\!R}^3}$. As it varies with a fixed origin, its tip traces out a unit 2-sphere in ${{\rm I\!R}^3}$. At a given fixed tip, ${{\bf q}(\eta_{{\bf x}{\bf y}},\,{\bf r})}$ defined in (\ref{three}) then simply acts like a complex number, because ${{\bf J}({\bf r})}$, with ${{\bf J}^2({\bf r})=-1}$, acts like an imaginary unit ${\pm\,i}$. As the angle ${\eta_{{\bf x}{\bf y}}}$ varies, ${{\bf q}(\eta_{{\bf x}{\bf y}},\,{\bf r})}$ thus traces out a circle, ${S^1}$. Since locally this is true at every point of ${S^2}$, locally ${S^3 = S^2\times S^1}$. In other words, locally, ${S^3}$ is simply ${S^2}$ worth of circles. Globally, however, in the topological sense, it has no cross-section \cite{Penrose}\cite{Ryder}. The best it can be viewed is as a principal U(1) bundle over ${S^2}$, with the points of its base space ${S^2}$ being the bivectors ${{\bf J}({\bf r})}$ \cite{Eguchi}. As in Eq.~(\ref{20}), the product of two such bivectors are in general non-pure quaternions and thus points of the bundle space ${S^3}$. The elements of ${S^3}$ are thus preimages of the points of the base space ${S^2}$. These preimages are 1-spheres, ${S^1}$, called Hopf circles, or Clifford parallels. Since these\break 1-spheres are the fibers of the bundle, they do not share a single point in common. As shown in Fig.~\ref{fig4}, each circle, ${S^1}$, threads through every other circle in the bundle, making them linked together in a highly nontrivial configuration~\cite{Lyons}.

It is these nontrivial M\"obius-like twists in the Hopf bundle of ${S^3}$ \cite{Penrose}, in conjunction with the spinorial sign changes discussed above, that are responsible for the origin and strength of the observed strong correlations (\ref{calcul}). Moreover, as we have proved in Section VIII of Ref.~\cite{Christian2014} and the Appendix \ref{A} below, in the context of EPR-Bohm type experiments these M\"obius-like twists in ${S^3}$ are equivalent to an algebraic expression for the conservation of total spin angular momentum. We therefore proceed to derive the strong correlations {\it directly} from the conservation of spin angular momentum \cite{Christian2014}. To this end, we require that the total spin of the singlet system emerging from the common source respects the condition
\begin{equation}
-\,{\bf L}({\bf s}_1,\,\lambda^k)\,+\,{\bf L}({\bf s}_2,\,\lambda^k)\,=\,0 \,\Longleftrightarrow\;\; {\bf L}({\bf s}_1,\,\lambda^k)\,=\,{\bf L}({\bf s}_2,\,\lambda^k) \,\Longleftrightarrow\;\; {\bf s}_1=\,{\bf s}_2\,\equiv\,{\bf s}\;
\;\;\,[\text{cf. Fig.~\ref{fig-1}}]. \label{56}
\end{equation}
Evidently, in the light of the product rule (\ref{50}) for the unit bivectors, the above condition is equivalent to the condition
\begin{equation}
{\bf L}({\bf s}_1,\,\lambda^k)\,{\bf L}({\bf s}_2,\,\lambda^k)=\left\{\,{\bf L}({\bf s},\,\lambda^k)\right\}^2=\,{\bf L}^2({\bf s},\,\lambda^k)=-1\,. \label{566}
\end{equation}
Note, however, that the limits ${\mathbf{s}_1\to \mathbf{a}}$ and ${\mathbf{s}_2\to \mathbf{b}}$ appearing in the definitions of the two measurement functions (\ref{53}) and (\ref{54}) are parts of the {\it independent} detection processes \cite{Christian}. These processes are {\it not} subject to the conservation law dictated by Eq.~(\ref{56}) or (\ref{566}), which remains valid only for the free evolution of the constituent spins \cite{Christian2014}. In fact, the detection processes describe purely local interactions of the spin bivectors with the detector bivectors, occurring at spacelike separated observation stations of Alice and Bob. Consequently, the expectation value of the simultaneous measurement outcomes ${{\mathscr A}({\bf a},\,{\lambda^k})=\pm1}$ and ${{\mathscr B}({\bf b},\,{\lambda^k})=\pm1}$, as limiting scalar points within ${S^3}$, works out as follows:
\begin{align}
{\cal E}_{\rm L.R.}({\bf a},\,{\bf b})\,&=\lim_{\,n\,\gg\,1}\left[\frac{1}{n}\sum_{k\,=\,1}^{n}\,{\mathscr A}({\bf a},\,{\lambda}^k)\;{\mathscr B}({\bf b},\,{\lambda}^k)\right] \label{57} \\
&=\lim_{\,n\,\gg\,1}\left[\frac{1}{n}\sum_{k\,=\,1}^{n}\,\bigg[\lim_{{\bf s}_1\,\rightarrow\,{\bf a}}\left\{\,-\,{\bf D}({\bf a})\,{\bf L}({\bf s}_1,\,\lambda^k)\right\}\bigg]\left[\lim_{{\bf s}_2\,\rightarrow\,{\bf b}}\left\{\,+\,{\bf L}({\bf s}_2,\,\lambda^k)\,{\bf D}({\bf b})\right\}\,\right]\right] \label{58}\\
&=\lim_{\,n\,\gg\,1}\left[\frac{1}{n}\sum_{k\,=\,1}^{n}\,\lim_{\substack{{\bf s}_1\,\rightarrow\,{\bf a} \\ {\bf s}_2\,\rightarrow\,{\bf b}}}\left\{\,-\,{\bf D}({\bf a})\,\right\}\,\left\{\,{\bf L}({\bf s}_1,\,\lambda^k)\,\,{\bf L}({\bf s}_2,\,\lambda^k)\,\right\}\,\left\{\,+\,{\bf D}({\bf b})\,\right\}\right] \label{59} \\
&=\lim_{\,n\,\gg\,1}\left[\frac{1}{n}\sum_{k\,=\,1}^{n}\,\lim_{\substack{{\bf s}_1\,\rightarrow\,{\bf a} \\ {\bf s}_2\,\rightarrow\,{\bf b}}}\left\{-\,\lambda^k\,{\bf L}({\bf a},\,\lambda^k)\right\}\,\left\{\,-1\,\right\}\,\left\{+\,\lambda^k\,{\bf L}({\bf b},\,\lambda^k)\right\}\right] \label{60}\\
&=\lim_{\,n\,\gg\,1}\left[\frac{1}{n}\sum_{k\,=\,1}^{n}\,\lim_{\substack{{\bf s}_1\,\rightarrow\,{\bf a} \\ {\bf s}_2\,\rightarrow\,{\bf b}}}\left\{\,+\,\left(\lambda^k\right)^2\,{\bf L}({\bf a},\,\lambda^k)\,\,{\bf L}({\bf b},\,\lambda^k)\right\}\right] \label{61} \\
&=\lim_{\,n\,\gg\,1}\left[\frac{1}{n}\sum_{k\,=\,1}^{n}\,{\bf L}({\bf a},\,\lambda^k)\,{\bf L}({\bf b},\,\lambda^k)\,\right]. \label{62}
\end{align}
In the above derivation, Eq.${\,}$(\ref{58}) follows from Eq.${\,}$(\ref{57}) by substituting the functions ${{\mathscr A}({\bf a},\,{\lambda^k})}$ and ${{\mathscr B}({\bf b},\,{\lambda^k})}$ from their definitions (\ref{53}) and (\ref{54}); Eq.${\,}$(\ref{59}) follows from Eq.${\,}$(\ref{58}) by using the ``product of limits equal to limits of product'' rule [which can be verified by recognizing that the same quaternion ${-\,{\bf D}({\bf a})\,{\bf L}({\bf a},\,\lambda^k)\,{\bf L}({\bf b},\,\lambda^k)\,{\bf D}({\bf b})}$ results from the limits in Eqs.${\,}$(\ref{58}) and (\ref{59})]; Eq.${\,}$(\ref{60}) follows from Eq.${\,}$(\ref{59}) by (i) using the relation (\ref{55}) [thus setting all bivectors in the spin bases], (ii) the associativity of the geometric product, and (iii) the conservation of spin angular momentum specified in Eq.${\,}$(\ref{566}); Eq.${\,}$(\ref{61}) follows from Eq.${\,}$(\ref{60}) by recalling that scalars such as ${\lambda^k}$ commute with the bivectors; and Eq.${\,}$(\ref{62}) follows from Eq.${\,}$(\ref{61}) by using the fact that ${\lambda^2 = +1}$, and by removing the superfluous limit operations.

The final sum in Eq.~(\ref{62}) can now be evaluated using Eqs.~(\ref{50}) and (\ref{55}), by recognizing that the observed spins in the right- and left-oriented ${S^3}$ satisfy the following geometrical relations (for a more detailed calculation see Ref.~\cite{Christian2014}):
\begin{equation}
{\bf L}({\bf a},\,{\lambda}^k=+1)\;{\bf L}({\bf b},\,{\lambda}^k=+1)\,=\,-\,{\bf a}\cdot{\bf b}\,-\,{\bf D}({\bf a}\times{\bf b}) \,=\,{\bf D}({\bf a})\;{\bf D}({\bf b}) \,=\,(\,+\,I\cdot{\bf a})(\,+\,I\cdot{\bf b}) \label{87}
\end{equation}
and
\begin{equation}
{\bf L}({\bf a},\,{\lambda}^k=-1)\;{\bf L}({\bf b},\,{\lambda}^k=-1)\,=\,-\,{\bf a}\cdot{\bf b}\,+\,{\bf D}({\bf a}\times{\bf b}) \,=\,-\,{\bf b}\cdot{\bf a}\,-\,{\bf D}({\bf b}\times{\bf a}) \,=\,{\bf D}({\bf b})\;{\bf D}({\bf a}) \,=\,(\,+\,I\cdot{\bf b})(\,+\,I\cdot{\bf a}). \label{88}
\end{equation}
In other words, when ${\lambda^k}$ happens to be equal to ${+1}$, ${{\bf L}({\bf a},\,{\lambda}^k)\;{\bf L}({\bf b},\,{\lambda}^k)=(\,+\,I\cdot{\bf a})(\,+\,I\cdot{\bf b})}$, and when ${\lambda^k}$ happens to be equal to ${-1}$, ${{\bf L}({\bf a},\,{\lambda}^k)\;{\bf L}({\bf b},\,{\lambda}^k)=(\,+\,I\cdot{\bf b})(\,+\,I\cdot{\bf a})}$. Consequently, the expected value in (\ref{62}) reduces at once to
\begin{equation}
{\cal E}_{\rm L.R.}({\bf a},\,{\bf b})\,=\,\frac{1}{2}(\,+\,I\cdot{\bf a})(\,+\,I\cdot{\bf b})\,+\,\frac{1}{2}(\,+\,I\cdot{\bf b})(\,+\,I\cdot{\bf a})\,
=\,-\,\frac{1}{2}\left\{{\bf a}{\bf b}\,+\,{\bf b}{\bf a}\right\}=\,-\,{\bf a}\cdot{\bf b}\,+\,0\,,\label{65b}
\end{equation}
because the handedness ${\lambda^k}$ of ${S^3}$ is a fair coin. Here the last equality follows from the definition of the inner product.

It is instructive to derive the above result somewhat differently to bring out the vital role played by the conservation of spin angular momentum (\ref{56}) in the derivation. To this end, note that for ${{\bf s}_1={\bf s}_2}$ the angle ${\eta_{{\bf u}{\bf v}}}$ in (\ref{38}) reduces to the angle ${\eta_{{\bf a}{\bf b}}}$ between the detector directions ${\bf a}$ and ${\bf b}$, and then, by using (\ref{null}), the product derived in (\ref{35}) tends to
\begin{align}
{\mathscr A}({\bf a},\,{\lambda^k})\,{\mathscr B}({\bf b},\,{\lambda^k})\,=\lim_{\substack{{\bf s}_1\,\rightarrow\,{\bf a} \\ {\bf s}_2\,\rightarrow\,{\bf b}}}\left\{\,-\,{\bf q}(\eta_{{\bf a}{\bf b}},\,{\bf r}_{0})\right\}\,
&=\lim_{\substack{{\bf s}_1\,\rightarrow\,{\bf a} \\ {\bf s}_2\,\rightarrow\,{\bf b}}}\left\{-\cos(\,\eta_{{\bf a}{\bf b}})-{\bf L}({\bf r}_{0},\,\lambda^k)\,\sin(\,\eta_{{\bf a}{\bf b}})\right\}, \notag \\
&=\,-\cos(\,\eta_{{\bf a}{\bf b}})-{\bf L}(\vec{\bf \,0},\,\lambda^k)\,\sin(\,\eta_{{\bf a}{\bf b}}), \label{54b}
\end{align}
where ${{\bf L}(\vec{\bf \,0},\,\lambda^k)}$ is a null bivector (just as a null vector is a vector without a magnitude or direction, a null bivector is a bivector without a magnitude or direction). Evidently, this tendency of ${{\mathscr A}{\mathscr B}}$ holds for each run of the experiment. To reflect this, in the following derivation of the correlations between the outcomes ${{\mathscr A}({\bf a},\,{\lambda^k})=\pm1}$ and ${{\mathscr B}({\bf b},\,{\lambda^k})=\pm1}$ we assume ${{\bf s}_1={\bf s}_2}$ from the outset during the free evaluation of the spins prior to their detections by Alice and Bob:
\begin{align}
{\cal E}_{\rm L.R.}({\bf a},\,{\bf b})\,
&=\lim_{\,n\,\gg\,1}\left[\frac{1}{n}\sum_{k\,=\,1}^{n}\,{\mathscr A}({\bf a},\,{\lambda}^k)\;{\mathscr B}({\bf b},\,{\lambda}^k)\right] \label{57a} \\
&=\lim_{\,n\,\gg\,1}\left[\frac{1}{n}\sum_{k\,=\,1}^{n}\,\left[\lim_{{\bf s}_1\,\rightarrow\,{\bf a}}\left\{\,+\,{\bf q}(\eta_{{\bf a}{\bf s}_1},\,{\bf r}_1)\right\}\right]\left[\lim_{{\bf s}_2\,\rightarrow\,{\bf b}}\left\{\,-\,{\bf q}(\eta_{{\bf s}_2{\bf b}},\,{\bf r}_2)\right\}\right]
\right] \label{58a}\\
&=\lim_{\,n\,\gg\,1}\left[\frac{1}{n}\sum_{k\,=\,1}^{n}\,\lim_{\substack{{\bf s}_1\,\rightarrow\,{\bf a} \\ {\bf s}_2\,\rightarrow\,{\bf b}}}\left\{\,-\,{\bf q}(\eta_{{\bf a}{\bf s}_1},\,{\bf r}_1)\,{\bf q}(\eta_{{\bf s}_2{\bf b}},\,{\bf r}_2)\right\}\right] \label{59a} \\
&=\lim_{\,n\,\gg\,1}\left[\frac{1}{n}\sum_{k\,=\,1}^{n}\,\lim_{\substack{{\bf s}_1\,\rightarrow\,{\bf a} \\ {\bf s}_2\,\rightarrow\,{\bf b}}}\left\{\,-\,{\bf q}(\eta_{{\bf a}{\bf b}},\,{\bf r}_{0})\right\}\right] \label{60a}\\
&=\lim_{\,n\,\gg\,1}\left[\frac{1}{n}\sum_{k\,=\,1}^{n}\,\{-\cos(\,\eta_{{\bf a}{\bf b}})-{\bf L}(\vec{\bf \,0},\,\lambda^k)\,\sin(\,\eta_{{\bf a}{\bf b}})\}\right] \label{61a} \\
&=\,-\cos(\,\eta_{{\bf a}{\bf b}})\,-\!\lim_{\,n\,\gg\,1}\left[\frac{1}{n}\sum_{k\,=\,1}^{n}\,{\bf L}(\vec{\bf \,0},\,\lambda^k)\,\sin(\,\eta_{{\bf a}{\bf b}})\,\right]\label{63a}\\
&=\,-\cos(\,\eta_{{\bf a}{\bf b}})\,-\!\lim_{\,n\,\gg\,1}\left[\frac{1}{n}\sum_{k\,=\,1}^{n}\,\lambda^k\,\right]{\bf D}(\vec{\bf \,0}\,)\,\sin(\,\eta_{{\bf a}{\bf b}}) \label{64a}\\
&=\,-\cos(\,\eta_{{\bf a}{\bf b}})\,-\,0\,. \label{65a}
\end{align}
In the above derivation, Eq.${\,}$(\ref{58a}) follows from Eq.${\,}$(\ref{57a}) by substituting the functions ${{\mathscr A}({\bf a},\,{\lambda^k})}$ and ${{\mathscr B}({\bf b},\,{\lambda^k})}$ from their definitions (\ref{53}) and (\ref{54}); Eq.${\,}$(\ref{59a}) follows from Eq.${\,}$(\ref{58a}) by using the ``product of limits equal to limits of product'' rule, as in the previous derivation of the correlations; Eq.${\,}$(\ref{60a}) follows from Eq.${\,}$(\ref{59a}) by multiplying the two quaternions in Eq.${\,}$(\ref{59a}), as carried out in Eq.${\,}$(\ref{35}), and by setting ${{\bf s}_1={\bf s}_2}$ in Eq.~(\ref{38}); Eq.${\,}$(\ref{61a}) follows from Eq.${\,}$(\ref{60a}) by performing the limit operations, as in Eq.${\,}$(\ref{54b}); Eq.${\,}$(\ref{63a}) follows from Eq.${\,}$(\ref{61a}) by noticing that ${\lambda^k}$ appears only in the second term of the Eq.${\,}$(\ref{61a}); Eq.${\,}$(\ref{64a}) follows from Eq.${\,}$(\ref{63a}) by using the relation (\ref{55}); and Eq.${\,}$(\ref{65a}) follows from Eq.${\,}$(\ref{64a}) as the scalar coefficient of the null bivector ${{\bf D}(\vec{\bf \,0}\,)}$ vanishes in the large-${n}$ limit because the handedness ${\lambda^k=\pm1}$ of ${S^3}$ is a fair coin. In fact, Eq.~(\ref{64a}) is redundant, because the $n$ null bivectors, ${{\bf L}(\vec{\bf \,0},\,\lambda^k)\sin(\,\eta_{{\bf a}{\bf b}})}$, appearing in Eq.~(\ref{63a}) are all {\it additive identities}. They add up giving a new combined null bivector, thereby reducing Eq.~(\ref{63a}) to Eq.~(\ref{65a}) at once.

Note that, apart from the initial or complete state ${\lambda^k}$, the only other assumption used in the derivations of the strong correlations (\ref{65a}) and (\ref{65b}) is that of the conservation of spin angular momentum, as specified in Eq.~(\ref{56}), or, equivalently, in Eq.~(\ref{566}). These two assumptions are both necessary and sufficient to dictate the singlet correlations
\begin{equation}
{\cal E}_{\rm L.R.}({\bf a},\,{\bf b})\,=\lim_{\,n\,\gg\,1}\left[\frac{1}{n}\sum_{k\,=\,1}^{n}\,{\mathscr A}({\bf a},\,{\lambda}^k)\;{\mathscr B}({\bf b},\,{\lambda}^k)\right]\!
=-\,{\bf a}\cdot{\bf b}. \label{abcos}
\end{equation}
Moreover, in the Appendix \ref{A} below we have demonstrated that the conservation of spin angular momentum is not an additional assumption but follows from the M\"obius-like twists in the very geometry of the quaternionic 3-sphere. This corroborates our hypothesis that singlet correlations are correlations amongst the points of a quaternionic 3-sphere.

Although the above analytical derivation of the singlet correlations speaks for itself, it can be further verified by an event-by-event numerical simulation. A code for such a numerical simulation is provided in the Appendix \ref{B} below.

\section{Concluding Remarks}

Our goal in this paper has been to explain the observed singlet correlations between a pair of entangled fermions as Dr.~Bertlmann’s socks type classical correlations \cite{Bell-1987}\cite{EPR}\cite{Bell-1990}. Because the full technical details of our explanation have been published in Refs.~\cite{Christian}, \cite{RSOS} and \cite{Christian2014}, the purpose of the present exposition is largely pedagogical. For this reason, in Section \ref{II} of this paper we began with a toy model of a fictitious one-sided world of M\"obius strip and showed that the strong correlations observed by the two-dimensional Alice and Bob living in this world are Dr.~Bertlmann's socks type classical correlations, despite the fact that the value of the product ${{\mathscr A}{\mathscr B}}$ of the local measurement results obtained by Alice and Bob does not remain fixed at ${-1}$ for nonparallel directions of their detectors. For some directions ${{\mathscr A}{\mathscr B}}$ fluctuates from ${-1}$ to ${+1}$ despite the correlations between their results being Dr.~Bertlmann's socks type correlations. Consequently, all four possible combinations of outcomes, ${+\,+}$, ${+\,-}$, ${-\,+}$, and ${-\,-}$, are observed by Alice and Bob.

The mathematical relation that facilitates similar explanation in our real, three-dimensional world is the following well known spinorial sign changes exhibited by quaternions when we use them to represent rotations in physical space:
\begin{equation}
{\bf q}(\eta_{{\bf x}{\bf y}}+\kappa\pi,\,{\bf r})\,=\,-\,{\bf q}(\eta_{{\bf x}{\bf y}},\,{\bf r})\,\;\;\text{for}\;\,\kappa=1,3,5,7,\dots, \tag{\ref{23}}
\end{equation}
where ${\eta_{{\bf x}{\bf y}}}$ is the angle between the bivectors ${{\bf J}({\bf x})}$ and ${{\bf J}({\bf y})}$ factorizing the quaternion ${{\bf q}(\eta_{{\bf x}{\bf y}},\,{\bf r})}$, with the unit vector ${{\bf r}\in{\rm I\!R}^3}$ representing its axis of rotation \cite{Christian}\cite{Christian2014}. However, Alice and Bob having observed all four possible combinations of measurement outcomes, ${{\mathscr A}{\mathscr B}=+\,+}$, ${+\,-}$, ${-\,+}$, and ${-\,-}$, is, by itself, not sufficient to account for the strong correlations between their results. Therefore, our central hypothesis in Refs.~\cite{Christian}, \cite{RSOS}, \cite{Christian2014} and \cite{disproof} has been that, instead of ${{\rm I\!R}^3}$, the physical space we live in should be modeled by a parallelizable 3-sphere, ${S^3}$, defined as a set of quaternions:
\begin{equation}
S^3:=\left\{\,{\bf q}(\eta_{{\bf x}{\bf y}},\,{\bf r}):=\cos(\,\eta_{{\bf x}{\bf y}})+{\bf J}({\bf r})\,\sin(\,\eta_{{\bf x}{\bf y}})\,
\Bigg|\;||\,{\bf q}(\eta_{{\bf x}{\bf y}},\,{\bf r})\,||^2=1\right\}. \tag{\ref{three}} 
\end{equation}
The strong correlations we observe in Nature between the measurement results of Alice and Bob are then accounted for as Dr.~Bertlmann's socks type classical correlations. 
This is possible because of the highly nontrivial geometrical and topological properties of ${S^3}$, despite it being only a three-dimensional space and a spatial part of a well known solution of Einstein's field equations of general relativity. In particular, the origin and strength of the strong correlations can be traced to the nontrivial twists in the Hopf bundle of ${S^3}$. Fortunately, these twists turn out to be directly related to\break the conservation of spin angular momentum in the EPR-Bohm type experiments, which can be expressed simply as
\begin{equation}
{\bf L}^2({\bf s},\,\lambda)=-1, \tag{\ref{566}} 
\end{equation}
where the bivector ${{\bf L}({\bf s},\,\lambda)}$ represents a spin. This condition dictates the strong correlations and guarantees that results of the experiments exist within ${S^3}$ before the experiments are performed, in line with Einstein's ideas of locality and realism \cite{EPR}. Conversely, removing this condition -- which amounts to reducing ${S^3}$ to ${{\rm I\!R}^3}$ by surgically removing a single\break point from ${S^3}$ -- reduces the strong correlations to perfect anti-correlation, such as that between Dr.~Bertlmann's socks.

\appendix

\section{From the Conservation of Spin to the Twists in the Hopf Bundle of $S^3$}\label{A}

\begin{figure}[t]
\vspace{-0.3cm}
\centering
\includegraphics[scale=1]{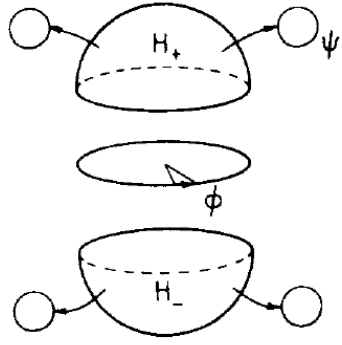}
\hrule
\caption{The Hopf bundle of ${S^3}$ showing two hemispherical neighborhoods ${{\rm H}_{\pm}}$ covering the base manifold ${S^2}$ \cite{Eguchi}. A fiber U(1) = ${S^1}$ parametrized by ${\psi}$ is attached to each point of ${{\rm H}_{\pm}}$.~The intersection of ${{\rm H}_{\pm}}$ at ${\theta\sim\frac{\pi}{2}}$ is a circular strip parametrized by ${\phi}$.}
\vspace{0.2cm}
\hrule
\label{fig7}
\end{figure}

In the Section VIII of Ref.~\cite{Christian2014} we have derived the conservation of spin from the twists in the Hopf bundle of ${S^3}$. In this appendix we derive the converse. To this end, we begin by multiplying both the numerator and denominator of the RHS of ${{\bf L}({\bf s}_1,\,\lambda^k)={\bf L}({\bf s}_2,\,\lambda^k)}$ from Eq.~(\ref{56}), with ${-{\bf D}({\bf b})}$ from the left and ${+{\bf D}({\bf b})}$ from the right, which gives
\begin{equation}
{\bf L}({\bf s}_1,\,\lambda^k)\,=\,\frac{\,-\,{\bf D}({\bf b})\,{\bf L}({\bf s}_2,\,\lambda^k)\,{\bf D}({\bf b})\,}{-\,{\bf D}({\bf b})\,{\bf D}({\bf b})\,}\,=\,-\,{\bf D}({\bf b})\,{\bf L}({\bf s}_2,\,\lambda^k)\,{\bf D}({\bf b}), \label{5}
\end{equation}
because ${{\bf D}^2({\bf b})=-1}$ for the unit bivector ${{\bf D}({\bf b})}$. Next,  multiplying both sides of (\ref{5}) from the left with ${-\,{\bf D}({\bf a})}$ gives
\begin{equation}
\big\{\!-{\bf D}({\bf a})\,{\bf L}({\bf s}_1,\,\lambda^k)\big\}=\big\{{\bf D}({\bf a})\,{\bf D}({\bf b})\big\}\big\{{\bf L}({\bf s}_2,\,\lambda^k)\,{\bf D}({\bf b})\big\}
\end{equation}
[cf. Eqs.~(\ref{53}) and (\ref{54})]. Using Eq.~(\ref{55}), ${{\bf D}({\bf n})=I\cdot{\bf n}}$ for any unit vector ${\bf n}$, and ${I^2=-1}$, this equality simplifies to
\begin{align}
\big\{\!-{\bf D}({\bf a})\,{\bf D}({\bf s}_1)\big\}&=\big\{\!-{\bf D}({\bf a})\,{\bf D}({\bf b})\big\}\big\{\!-{\bf D}({\bf s}_2)\,{\bf D}({\bf b})\big\} \\
\Longleftrightarrow\;\;({\bf a}\,{\bf s}_1)&=({\bf a}\,{\bf b})\,({\bf s}_2\,{\bf b}).
\end{align}
Using ${{\bf s}_{1}={\bf s}_{2}={\bf s}}$ from the conservation of spin angular momentum stated in (\ref{56}), the above relation is equivalent to
\begin{equation}
\exp\left[{\bf J}({\bf r}_{{\bf a}{\bf s}})\,\eta_{{\bf a}{\bf s}}\right]=\exp\left[{\bf J}({\bf r}_{{\bf a}{\bf b}})\,\eta_{{\bf a}{\bf b}}\right]\,\exp\left[{\bf J}({\bf r}_{{\bf s}{\bf b}})\,\eta_{{\bf s}{\bf b}}\right], \label{o}
\end{equation}
where each of the three bivectors, ${{\bf J}({\bf r}_{{\bf a}{\bf s}})=\frac{{\bf a}\wedge{\bf s}}{||{\bf a}\wedge{\bf s}||}}$, ${{\bf J}({\bf r}_{{\bf a}{\bf b}})=\frac{{\bf a}\wedge{\bf b}}{||{\bf a}\wedge{\bf b}||}}$, and ${{\bf J}({\bf r}_{{\bf s}{\bf b}})=\frac{{\bf s}\wedge{\bf b}}{||{\bf s}\wedge{\bf b}||}}$, function just like the familiar imaginary scalar unit $i$, and ${\eta_{{{\bf a}}{{{\bf s}}}}}$, ${\eta_{{{\bf a}}{{{\bf b}}}}}$, and ${\eta_{{\bf s}{\bf b}}}$ are angles between the vectors ${{\bf a}}$ and ${{{\bf s}}}$, ${{{\bf a}}}$ and ${{\bf b}}$, and ${{\bf s}}$ and ${{{\bf b}}}$, respectively.
Next, recall that the twists in the Hopf bundle of ${S^3}$ can be quantified by the exponential relation analogous to (\ref{o}):
\begin{equation}
e^{i\psi_-}\,=\,e^{i\phi}\,e^{i\psi_+}\,, \label{p}
\end{equation}
where, as shown in Fig.~\ref{fig7} \cite{Eguchi}, ${e^{i\psi_-}}$ and ${e^{i\psi_+}}$, respectively, are the U(1) fiber coordinates above the two hemispheres ${{\rm H}_-}$ and ${{\rm H}_+}$ of the base space ${S^2}$, with spherical coordinates ${(0\leqslant\theta < \pi,\;0\leqslant\phi < 2\pi)}$; ${\phi}$ is the angle parameterizing a thin strip ${{\rm H}_-\cap {\rm H}_+}$ around the equator of ${S^2}$ [${\theta\sim\frac{\pi}{2}}$]; and ${e^{i\phi}}$ is the transition function that glues the two sections ${{\rm H}_-}$ and ${{\rm H}_+}$ together, thus constituting the 3-sphere. It is evident from the above relation that the fibers match perfectly at the angle ${\phi=0}$ (modulo ${2\pi}$), but differ from each other at all intermediate angles ${\phi}$. For example, ${e^{i\psi_-}}$ and ${e^{i\psi_+}}$ differ by a minus sign at the angle ${\phi=\pi}$. The relations (\ref{o}) and (\ref{p}) can be identified, provided we identify the angles\break ${\eta_{{{\bf a}}{{{\bf s}}}}}$ and ${\eta_{{{\bf s}}{{{\bf b}}}}}$ between ${{\bf a}}$ and ${{{\bf s}}}$ and ${{{\bf s}}}$ and ${{\bf b}}$ with the fibers ${\psi_-}$ and ${\psi_+\,}$, and the angle ${\eta_{{\bf a}{\bf b}}}$ between ${{\bf a}}$ and ${{{\bf b}}}$ with the generator of the transition function ${e^{i\phi}}$ on the equator of ${S^2}$. We have thus shown the equivalence of the twists in the Hopf bundle of ${S^3}$ to the algebraic expression ${{\bf L}^2({\bf s},\,\lambda^k)=\,-1}$ for the conservation of zero spin angular momentum~\cite{Christian2014}.
\section{Event-by-Event Numerical Simulation of the singlet Correlations}\label{B}

The following code for the event-by-event simulation was written in collaboration with Carl F. Diether III \cite{Diether}. It is\break a 3D modification of the original code written by Albert Jan Wonnink to verify this quaternionic 3-sphere model \cite{Wonnink}.

\begin{alltt}
//Adaptation of Albert Jan Wonnink's GAViewer code for the \(S\sp{3}\) model of the singlet correlations

function getRandomLambda() 
\{
   if( rand()>0.5) \{return 1;\} else \{return -1;\}
\}
function getRandomUnitVector() //unit vector uniformly distributed over \(S\sp{2}\)
                               //\url{http://mathworld.wolfram.com/HyperspherePointPicking.html}
\{
   v=randGaussStd()*e1+randGaussStd()*e2+ randGaussStd()*e3;   //three-dimensional vectors
   return normalize(v);
\}
   batch test()
\{
   set_window_title("Test of 3D S^3 GA Model for the 2-particle singlet correlations");
   default_model(p3ga);                    //choice of the model in GAViewer 
   N=50000;                                //number of iterations (or trials)
   I=e1^e2^e3;                             //the fundamental trivector of GA
   ss=0;
   t=0;
   u=0;
   for(nn=0;nn<N;nn=nn+1)                  //perform the EPR-Bohm experiment N times
   \{
          a=getRandomUnitVector();
          Da=I a;                          //detector bivector of Alice as in Eq.\(\,\)(\ref{53})
          b=getRandomUnitVector();
          Db=I b;                          //detector bivector of Bob as in Eq.\(\,\)(\ref{54})
          s1=getRandomUnitVector();
          s2=s1;                           //conservation of zero spin as in Eq.\(\,\)(\ref{56})
          Ls1=I s1;                        //bivector representing the spin of particle 1
          Ls2=I s2;                        //bivector representing the spin of particle 2
          lambda=getRandomLambda();        //\(\lambda\) is a fair coin providing the \(\pm1\) choice 
          A=(-Da*lambda*Ls1);              //Alice's measurement function as in Eq.\(\,\)(\ref{53})
          B=(lambda*Ls2*Db);               //Bob's measurement function as in Eq.\(\,\)(\ref{54})
          //Note that the limits on A and B are not required because the conservation of
          //spin, s2=s1, is imposed, which reduces the product -Ls1*Ls2 in A*B to unity. 
          q=0;
          if(lambda==1) \{q=A B;\} else \{q=B A;\} //shuffles the alternative orientations of \(S\sp{3}\)
          ss=ss+q;
          phi_a=atan2(scalar(Da/(e3^e1)), scalar(Da/(e2^e3))); //gets azimuthal angle for \(\!\!\!\bf a\)
          phi_b=atan2(scalar(Db/(e2^e3)), scalar(Db/(e3^e1))); //gets azimuthal angle for \(\!\!\!\bf b\)
          neg_adotb=-(a.b);
          print(neg_adotb,"f" );               //outputs \(-\!\!{\bf a}\cdot\!\!\!{\bf b}\) event by event
          if(phi_a*phi_b>0) \{eta_ab=acos(a.b)*180/pi;\} else \{eta_ab=-acos(a.b)*180/pi+360;\}
          print(eta_ab, "f");                  //outputs the angles \(\eta\sb{\!\!\!\!{\bf a}\!\!\!\!{\bf b}}\) event by event
          print(correlation=scalar(q), "f");   //outputs correlations; see Ref.\(\,\)\cite{Diether} \& Fig.\(\,\)\ref{fig6}
          t=t+A;
          u=u+B;
      \}
      mean=ss/N;
      print(mean, "f");    //shows the vanishing of the non-scalar part
      aveA=t/N;
      print(aveA, "f");    //verifies that individual average < A > = 0
      aveB=u/N;
      print(aveB, "f");    //verifies that individual average < B > = 0
      prompt();
\}
\end{alltt}
\begin{figure}[t]
\hrule
\scalebox{0.75}{
\begin{pspicture}(4.5,-1.0)(5.0,5.9)
\psset{xunit=0.5mm,yunit=4cm}
\psaxes[axesstyle=frame,linewidth=0.01mm,tickstyle=full,ticksize=0pt,dx=90\psxunit,Dx=180,dy=1
\psyunit,Dy=+2,Oy=-1](0,0)(180,1.0)
\psline[linewidth=0.2mm,arrowinset=0.3,arrowsize=2pt 3,arrowlength=2]{->}(0,0.5)(190,0.5)
\psline[linewidth=0.2mm]{-}(45,0)(45,1)
\psline[linewidth=0.2mm]{-}(90,0)(90,1)
\psline[linewidth=0.2mm]{-}(135,0)(135,1)
\psline[linewidth=0.2mm,arrowinset=0.3,arrowsize=2pt 3,arrowlength=2]{->}(0,0)(0,1.2)
\psline[linewidth=0.35mm,linestyle=dashed,linecolor=gray]{-}(0,0)(90,1)
\psline[linewidth=0.35mm,linestyle=dashed,linecolor=gray]{-}(90,1)(180,0)
\put(2.1,-0.38){${90}$}
\put(6.5,-0.38){${270}$}
\put(-0.63,3.92){${+}$}
\put(-0.9,5.0){{\large ${{\cal E}^{\rm EPR}_{{\!}_{L.R.}}}$}${({\bf a},\,{\bf b})}$}
\put(-0.38,1.93){${0}$}
\put(9.65,1.95){\large ${\eta_{{\bf a}{\bf b}}}$}
\psplot[linewidth=0.35mm,linecolor=black]{0.0}{180}{x dup cos exch cos mul 1.0 mul neg 1 add}
\end{pspicture}}
\hrule
\caption{Plot of event-by-event numerical simulation of the singlet correlations predicted by the ${S^3}$ model \cite{Diether}. The x-axis depicts\break the angle in degrees between the vectors ${\bf a}$ and ${\bf b}$ chosen by Alice and Bob and the y-axis depicts the value of the correlations. The dotted straight lines represent the well known classical prediction of correlations described in Eq.~(\ref{dismee}) (see also Ref.~\cite{Peres}).}
\vspace{0.3cm}
\label{fig6}
\hrule
\end{figure}
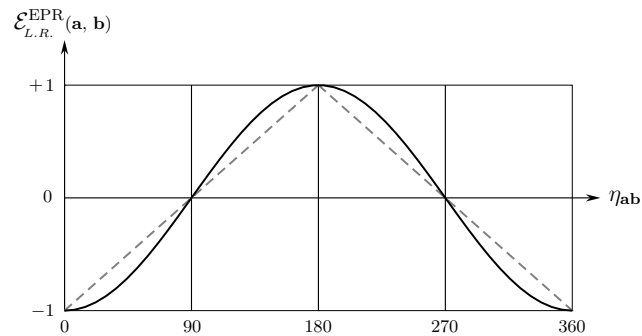
The graph generated by this simulation is shown in Fig.~\ref{fig6}. It is evident from it that the predictions of ${S^3}$ model match\break exactly with those of quantum mechanics ({\it i.e.}, with the negative cosine curve), despite the model being local-realistic.

\baselineskip 11pt

\acknowledgments
I am grateful to Carl F. Diether III for his help with simulating the singlet correlations presented in this paper. The code has been adapted from the original code written by Albert Jan Wonnink for simulating the 3-sphere model.

\end{document}